\documentclass[useAMS,usenatbib]{mn2e}
\usepackage{epsfig}
\usepackage{color}
\usepackage{natbib}
\usepackage{deluxetable}
\usepackage{amssymb}
\usepackage{rotating}
\usepackage{graphicx}
\usepackage{subfig}
\usepackage{IEEEtrantools}
\usepackage[colorlinks=true,citecolor=blue]{hyperref}
\title[Weak emission line Quasars]{Multi-epoch intra-night optical monitoring of 8 radio-quiet BL Lac candidates} 

\author[Kumar et al.] {{P. Kumar$^{1}$\thanks{E-mail: parveen@aries.res.in (PK)},
   Gopal-Krishna$^{2}$, C. S. Stalin$^{3}$, H. Chand$^{1}$, R. Srianand$^{4}$}, P. Petitjean$^{5}$ \\ $^{1}$Aryabhatta Research
  Institute of Observational Sciences (ARIES), Manora Peak, Nainital,
  263002 India\\ $^{2}$Centre for Excellence in Basic Sciences (CEBS), University of Mumbai campus (Kalina), Mumbai 400098, 
  India\\ $^{3}$Indian Institute of Astrophysics, Block II, Koramangala, Bangalore-560034, India\\ $^{4}$Inter-University 
  Centre for Astronomy and Astrophysics (IUCAA), Postbag 4, Ganeshkhind, Pune 411 007, India\\ $^{5}$UPMC-CNRS, UMR7095, Institut 
  d$^\prime$Astrophysique de Paris, F$-$75014 Paris, France\\}

\begin{document}
\date{Accepted ---. Received ---; in original form ---}

\pagerange{\pageref{firstpage}--\pageref{lastpage}} \pubyear{2009}

\maketitle

\label{firstpage}
\begin{abstract}

For a new sample of $8$ weak-line-quasars (WLQs) we report a sensitive
search { in $20$ intranight monitoring sessions}, for blazar-like
optical flux variations on hour-like and longer time scale
(day/month/year-like). The sample consists exclusively of the WLQs
that are not radio-loud and have either been classified as `radio-weak
probable BL Lac candidates' and/or are known to have exhibited at
least one episode of large, blazar-like optical variability. Whereas
only a hint of intra-night variability is seen { for two of these
  WLQs, J104833.5$+$620305.0($z$ = 0.219) and J133219.6$+$622715.9
  ($z$ = 3.15)}, statistically significant { inter-night} variability
{ at a few per cent level} is detected for three of the sources,
including the radio-intermediate WLQ J133219.6$+$622715.9 ($z$ = 3.15)
and the well known bona-fide radio-quiet WLQs J121221.5$+$534128.0
($z$ = 3.10) and WLQ J153259.9$-$003944.1 (z = 4.62). In the
rest-frame, this variability is intra-day and in the far-UV band. On
the time scale of a decade, we find for three of the WLQs large
brightness changes, amounting to 1.655$\pm$0.009, 0.163$\pm$0.010 and
0.144$\pm$0.018 mag, for J104833.5$+$620305.0, J123743.1$+$630144.9
and J232428.4$+$144324.4, respectively. Whereas the latter two are
confirmed radio-quiet WLQs, the extragalactic nature of
J104833.5$+$620305.0 remains to be well established, thanks to the
absence of any feature(s) in its available optical spectra. The
present study forms a part of our ongoing campaign of intranight
optical monitoring of radio quiet weak-line quasars, in order to
improve the understanding of this enigmatic class of Active Galactic
Nuclei and to look among them for a possible tiny, elusive population
of radio-quiet BL Lacs.
    
\end{abstract}


\section{Introduction}

Being among the most active extragalactic sources, the
radio-to-optical/UV observations of BL Lacs are marked by strong and
rapid variability of both continuum and polarized emission, commonly
attributed to the dominance of a relativistically boosted jet of
synchrotron radiation, emanating from the core of the host
galaxy~\citep[e.g.,][]{Blandford1978PhyS...17..265B}. Possibly, the
same jet dominance effectively contributes to the weak appearance (or,
non-detection) of the emission lines in the optical/UV spectrum, which
is a defining characteristic of the BL Lac objects~\citep[e.g.,][and
  references
  therein]{Angel1980ARA&A..18..321A,Begelman1984RvMP...56..255B,
  Urry1995PASP..107..803U,Antonucci2012A&AT...27..557A}.  The
availability of spectra of thousands of faint, star-like objects
covered in the large imaging/spectroscopic surveys, such as the Sloan
Digital Sky Survey~\citep[SDSS;
  e.g.,][]{York2000AJ....120.1579Y,Schneider2005AJ....130..367S} and
the Two-Degree Field QSO Redshift
Survey~\citep[2QZ;][]{Boyle2000MNRAS.317.1014B,Croom2001AAONw..97....3C}
paved the way for making systematic optical searches for BL Lac
objects, which had been previously found mainly by an optical
follow-up of the candidates first detected in the radio or X-ray
bands~\citep[e.g.,][and references
  therein]{1991ApJ...374..431S,Stocke1991ApJS...76..813S,Perlman1996ApJS..104..251P,
  Padovani2007ApJ...662..182P,Anderson2007AJ....133..313A}. Although,
on a limited scale, optical variability on month/year-like time scales
has also been employed as a technique for finding AGN
candidates~\citep[e.g.,][and references
  therein]{Villforth2010ApJ...723..737V,Bauer2009ApJ...705...46B}, the
big boost in this direction came with the advent of massive optical
surveys. Indeed, these optical surveys began to reveal in large
numbers distant, high-luminosity objects which qualified as potential
BL Lacs by virtue of their stellar appearance and an essentially
featureless optical spectrum. These objects were termed
`weak-line-quasars' (WLQs)~\citep[e.g.,][]
{Fan1999ApJ...526L..57F,Anderson2001AJ....122..503A,Londish2002MNRAS.334..941L,
  Collinge2005AJ....129.2542C,Schneider2005AJ....130..367S,Fan2006AJ....131.1203F,
  Londish2007MNRAS.374..556L,Plotkin2008AJ....135.2453P,DiamondStanic2009ApJ...699..782D}.
It was even suggested that the WLQs could be the long sought high-z
counterparts of BL Lacs~\citep[e.g.,][]
{Shemmer2006ApJ...644...86S,Stocke2001ASPC..227..184S}. However, the
radio and X-ray outputs as well as the strength and variability of
both continuum and polarized emission of most WLQs are generally
modest, in a striking contrast to BL Lac objects, putting a question
mark on their BL Lac classification
(e.g.,~\citet{Meusinger2014A&A...568A.114M} and references
therein;~\citet{Plotkin2015ApJ...805..123P};~\citet{Luo2015ApJ...805..122L};
see below).
 
The combination of huge size and depth of the optical surveys has also
revived interest in the old question whether a rare population of BL
Lacs exists which is radio-quiet, in analogy to the radio-quiet
quasars (radio-quietness is usually parameterised by the ratio of
radio and optical flux densities, R{\footnote{Radio-loudness is
    usually parametrized by the ratio (R) of flux densities at 5 GHz
    and at 4400\AA~in the rest-frame, and R $<$ 10 for radio-quiet
    quasars~\citep[e.g. see, ][]{Kellermann1989AJ.....98.1195K}.}.
  Attempts to discover radio-quiet BL Lacs, previously triggered by
  the large X-ray
  surveys~\citep[e.g.,][]{Chanan1982ApJ...261L..31C,Stocke1991ApJS...76..813S}
  found little substantiation from the follow-up optical polarimetry
  of the BL Lac candidates gleaned from the X-ray
  surveys~\citep[e.g.,][]{Borra1984ApJ...276..449B,Jannuzi1993ApJ...404..100J}.
  A similar lack of confirmation emerged from the polarimetric
  follow-up of the radio-quiet BL Lac candidates extracted from the
  various optical surveys including the massive surveys mentioned
  above~\citep[see, e.g.,][and references therein]
  {Impey1982MNRAS.201..849I,Stocke1990ApJ...348..141S,Smith2007ApJ...663..118S,
    Heidt2011A&A...529A.162H}.  Likewise, the broad-band spectral
  measurements and optical monitoring to detect continuum variability
  on month/year-like time scales have shown that radio-quiet WLQs
  (RQWLQs) are mostly distinct from BL Lacs
  ~\citep[e.g.,][]{DiamondStanic2009ApJ...699..782D,Lane2011ApJ...743..163L,
    Wu2012ApJ...747...10W,Meusinger2014A&A...568A.114M,Plotkin2015ApJ...805..123P,
    Kugler2015MNRAS.451.3385K}. A similar inference is reached when
  X-ray properties of high-z WLQs are
  examined~\citep{Shemmer2009ApJ...696..580S}.

A number of proposals have been put forward to explain the
uncharacteristically weak line emission from WLQs. In one scheme, the
weakness has been attributed to a deficit of gas in the broad
emission-line region (BELR) of WLQs, either because the quasar
activity has begun only
recently~\citep{Hryniewicz2010MNRAS.404.2028H,Murray1997ApJ...474...91M},
or due to some other possible
cause~\citep[e.g.,][]{Shemmer2010ApJ...722L.152S}. A competing
explanation ~\citep[e.g.,][]{Nikolajuk2012MNRAS.420.2518N} argues that
the BELR in WLQs has a very low covering factor against the flux of
ionizing photons incident from the accretion disk. Possibly, this
situation could arise, for instance, due to a shielding gas layer (or,
patchy shielding gas) between the hot accretion disk and the BELR,
which attenuates the flux of ionizing photons emerging from the
accretion disk~\citep{Lane2011ApJ...743..163L,
  Wu2011ApJ...736...28W,Wu2012ApJ...747...10W};
also,~\citep{Luo2015ApJ...805..122L}.  Alternative explanations
implicate the physics of disk accretion onto the central super-massive
black-hole~\citep[e.g.,][]
{Laor2011MNRAS.417..681L,Leighly2007ApJ...663..103L,Nicastro2003ApJ...589L..13N}.
While all these theoretical scenarios are debated, substantial
evidence has accumulated showing that the central engines of at least
some radio-quiet quasars `RQQs' (including their weak line
counterparts, the RQWLQs) are in fact capable of ejecting relativistic
jets which are mostly feeble and extended on just the parsec
scale. This inference is rooted in the radio continuum imaging
including VLBI and flux monitoring campaigns targeted at
RQQs~\citep[e.g.,][]
{Blundell1998MNRAS.299..165B,Barvainis2005ApJ...618..108B,Caccianiga2001MNRAS.328..867C,Ulvestad2005ApJ...621..123U,Falcke1996ApJ...473L..13F,Leipski2006A&A...455..161L,Herrera2016A&A...589L...2H}.
It is also interesting to recall the striking discovery of an
intermittently present, relativistically beamed nuclear jet of radio
and X-ray
emission~\citep{Blundell2003ApJ...591L.103B,Gallo2006MNRAS.365..960G}
in the $z$ = 0.94 WLQ PG 1407+265 (J140923.9+261821.1;
~\citep{McDowell1995ApJ...450..585M}, which has been classified as
`radio-quiet'. Likewise, the possibility of a relativistic jet in the
{$z$ = 4.62 RQWLQ J153259.96$-$003944.1} has been inferred on the
basis of its strong optical
variability~\citep{Stalin2005MNRAS.359.1022S}. However, the
significance and importance of these findings remains to be
consolidated via polarimetric monitoring of the RQWLQs short-listed as
radio-quiet BL Lac candidates, since a high and variable polarization
is the hallmark of bona-fide BL Lacs. A robust finding of even a
minuscule population of radio-quiet BL Lacs, which could well be
lurking in samples of RQWLQs~\citep[e.g.,
  see,][]{Collinge2005AJ....129.2542C,Shemmer2006ApJ...644...86S,
  Lane2011ApJ...743..163L,Wu2012ApJ...747...10W}, would provide an
important new insight into the nature of AGN jets. For instance, a
remarkable ramification would be that in some rare relativistic jets,
either the synchrotron { radio} emission is abnormally weak
intrinsically, or the relativistic Doppler boost somehow is
uncharacteristically ineffective in the radio domain than at optical
wavelengths. It is, however, clear that if at all such exotic objects
exist, they must be extremely rare, as suggested by the
afore-mentioned studies of WLQs employing a number of basic
diagnostics, such as the multi-wavelength spectral/polarimetric
measurements, and optical flux variability, all of which seem to favor
the notion that at least an overwhelming majority of RQWLQs is
different from BL Lacs, whether radio or X-ray selected
~\citep[e.g.,][]{DiamondStanic2009ApJ...699..782D,Shemmer2009ApJ...696..580S,Lane2011ApJ...743..163L,
  Kuegler2014A&A...569A..95K,Meusinger2014A&A...568A.114M,Plotkin2015ApJ...805..123P}.
This inference about rarity of jet emission in RQWLQs is further
  supported by the frequently observed deficit of X-ray emission,
  though it is somewhat unclear how much of it is caused due to
  absorption~\citep[see, e.g.,][]{Luo2015ApJ...805..122L}.

In a parallel attempt aimed at finding radio-quiet BL Lac candidates,
we initiated a program to search for { rapid} optical variability
(on hour-like time scale) in a few dozen RQWLQs which had shown some
signatures of potential radio-quiet BL Lacs. The candidates were
selected from the catalogs of WLQs published by
~\citet{Plotkin2010ApJ...721..562P} and
~\citet{Meusinger2014A&A...568A.114M}. The results obtained so far
under this program to search for intra-night optical variability
(INOV) in $25$ radio-quiet BL Lac candidates, monitored by us in $48$
sessions of minimum $3$ hour duration, have been reported in $4$
papers (~\citet{Gopal2013MNRAS.430.1302G}: Paper
I,~\citet{Chand2014MNRAS.441..726C}: Paper
II,~\citet{Kumar2015MNRAS.448.1463K}: Paper
III,~\citet{Kumar2016MNRAS.461..666K}: Paper IV). Only two events of
INOV were detected, out of the 48 monitoring sessions, corresponding
to an INOV duty cycle of $\sim$ 3$\%$, which is much lower than that
known for BL Lacs~\citep[e.g.,][]{Goyal2013MNRAS.435.1300G}. However,
on both occasions the observed INOV had a large amplitude ( $>$ 10\%),
hinting at a blazar-like behavior for the two RQWLQs, namely,
J090843.25$+$285229.8 (Paper II) and J140710.26$+$241853.6 (Paper
III). Conceivably, the inferred smaller duty cycle could be due to the
lack of a matching sensitivity for monitoring WLQs which are mostly
fainter in comparison to the classical BL Lacs that have typically
been covered in intranight monitoring programs (e.g., Papers III and
IV). In this paper, we present results from $20$ sessions of
intranight optical monitoring (using 1 to 2$-$meter class telescopes),
covering a new set of $8$ WLQs which are not radio-loud
(Table~\ref{tab:source_info}). Note that all these objects,
  excepting J140923.9$+$261821.1, had also been monitored under a
  separate program nearly a decade ago by some of the present authors
  and those independently taken unpublished observations have been
  included in the present work.

\section{The Sample selection}
Six out of the $8$ WLQs in the present sample are drawn from the list
of $27$ ``Potential radio-weak BL Lac candidates", derived by
~\citet{Collinge2005AJ....129.2542C}) from the SDSS
DR2~\citep{Abazajian2004AJ....128..502A} and the 2DF
survey~\citep{Boyle2000MNRAS.317.1014B}, { by imposing the criteria
  of radio-quietness ~\citep[Table~\ref{tab:source_info}; see
    also][]{Kellermann1989AJ.....98.1195K}, a nearly featureless
  optical spectrum (i.e., no line with an EW $>$ 5\AA ~in the
  rest-frame), lack of proper motion above 2.5$\sigma$ threshold and a
  broad-band color criterion ($g - r \ge 0.35$ or $r - i \ge
  0.13$)~\citep{Collinge2005AJ....129.2542C} which is} expected to
minimize contamination by Galactic stars, such as DC white
dwarfs. Application of some additional selection criteria consistent
with BL Lacs led them to the list of $27$ sources. These additional
criteria are: (i) absence of stellar absorption features at zero
redshift, (ii) a spectrum inconsistent with the stellar template, or
combination of galaxy Eigen spectra,{ (iii) the strength of Ca II H
  \& K break being $<$ $0.4$. Out of the $27$ candidate radio-quiet
BL Lacs, we extracted a representative set of six without any bias
towards optical variability, or redshift, such that our target
selection was primarily dictated by the availability of the telescope
time, scheduling considerations and the availability of at least
  two, but usually more, comparison stars on each CCD frame, within
  about $1$ mag of the target. 

An independent check was made with the catalog of
~\citet{Monet2003AJ....125..984M}, to ensure that all our sources are
consistent with having zero proper
motion(Table~\ref{tab:source_info}). { The remaining two WLQs in our
  sample are PG 1407$+$265 ($z$ = 0.94) and J153259.96$-$003944.1 ({
    $z$ = 4.62}). Both these well known radio-quiet WLQs are reported
  to have shown an episode of blazar-like strong optical flaring
  (Section 1).} The basic properties of our set of $8$ WLQs are
summarized in Table~\ref{tab:source_info}.

\begin{table*}
\begin{minipage}{500mm}
{
\caption{Basic parameters of the present sample of $8$ WLQs
\label{tab:source_info}}
\begin{tabular}{lc cccccccc cr}
\hline
\multicolumn{1}{l}{Name} & R.A.(J2000) & DEC.(J2000)                       &{ r (mag)} &   z{\footnote{Reference is
~\citet{Hewett2010MNRAS.405.2302H} for $^{*}$; SDSS catalog for $^{\dagger}$; ~\citet{McDowell1995ApJ...450..585M} for J140923.9+261821.1.
 and\\~\citet{Fan1999ApJ...526L..57F} for J153259.9$-$003944.1}}& 
Proper Motion{\footnote{~\citet{Monet2003AJ....125..984M}.}} & $\alpha_{r-o}${\footnote {The radio-optical spectral index $\alpha_{r-o}$ is usually taken to be less than 0.21~\citep[e.g.,][]{Luo2015ApJ...805..122L} for radio-quiet\\ quasars having radio-loudness parameter $R<10$,  where R is the ratio of flux densities at 5 GHz and at $4400\AA$ in rest frame\\~\citep[e.g.,][]{Kellermann1989AJ.....98.1195K}. Reference for $\alpha_{r-o}$ is {~\citet{Collinge2005AJ....129.2542C} for $^{*}$; ~\citet{Plotkin2010AJ....139..390P} for $\ddagger$;~\citet{Fan1999ApJ...526L..57F} for $\$$\\ and 
~\citet{Kellermann1989AJ.....98.1195K}} for $\divideontimes$.}}&Telescope\\
         & (h m s)       &($ ^{\circ}$ $ ^{\prime}$ $ ^{\prime\prime }$) &  &     &(mas/yr)\\        
 (1)     &(2)             &(3)                             &(4)     &(5)&(6)&(7)&(8)\\
\hline
\multicolumn{5}{l}{}\\

J104833.5$+$620305.0 {\footnote {The extragalctic nature of this source is uncertain (e.g., see, Section 5.1) as it has been classified\\ in the SDSS as a star.~\citet{Richards2009ApJS..180...67R} give its photometric redshift as $1.54$.}} & 10:48:33.56 & $+$62:03:05.0  &19.86 &$0.219$ $\pm0.178$$^{\dagger}$&$14.14\pm7.07$&$<0.27^{\star}$&HCT\\
J121221.5$+$534128.0 & 12:12:21.56 & $+$53:41:28.0  &18.63 &$3.0976$$\pm0.0003$$^{*}$&$0$&$<0.03^{\star}$&OHP\\
J123743.1$+$630144.9 & 12:37:43.14 & $+$63:01:44.9  &19.00 &$3.4278$$\pm0.0023$$^{*}$&$0$&$<0.04^{\star}$&OHP\\
J124225.3$+$642919.0 & 12:42:25.39 & $+$64:29:19.0  &15.69 &$0.04247$$\pm0.00002$$^{\dagger}$&$0$&$<-0.06^{\ddagger}$&OHP\\
J133219.6$+$622715.9 & 13:32:19.65 & $+$62:27:15.9  &19.19 &$3.1514$$\pm0.0023$$^{*}$&$8.94\pm3.61$ &$0.19^{\star}$&HCT\\
J140923.9$+$261821.1 & 14:09:23.91 & $+$26:18:21.1  &15.74 &$0.94\pm0.02$&$6.32\pm4.472$&$0.10^{\divideontimes}$&DFOT\\
J153259.9$-$003944.1 & 15:32:59.96 & $-$00:39:44.1  &21.16 &$4.62\pm0.04$&0&$<0.02^{\$}$& HCT\\
J232428.4$+$144324.4 & 23:24:28.43 & $+$14:43:24.4  &18.79&$1.4172$$\pm0.0006$$^{*}$&$0$&$<0.01^{\ddagger}$&HCT\\
          
\hline
\end{tabular}

}
\end{minipage}
\end{table*}
\section{Observation and reduction}
The photometric observations were carried out in the R-band using the
$2.0-$m Himalayan Chandra Telescope (HCT) of the Indian Astronomical
Observatory (IAO) at Hanle~\citep{Prabhu2010ASInC...1..193P}, the
$1.3-$m Devasthal Fast Optical Telescope (DFOT) located near Nainital
(India)~\citep{Sagar2011Csi...101...8.25} and the $1.2-$m telescope of
the Observatoire de Haute-Provence (OHP), France{\footnote
  {http://www.obs-hp.fr/}}.

The $2.0-$m HCT has a Ritchey-Chretien (RC) design with a f/9 beam at
the Cassegrain focus. The detector used is a cryogenically cooled
2048$\times$ 4096 chip, having a pixel size of $15$ micron and a plate
scale of 0.29 arcsec per pixel which covers an area of $\sim$ 10
arcmin on the sky.  The readout noise of the CCD is $4.8$ e-/pixel the
gain is $1.22$ e-/Analog to Digital Unit (ADU).

The $1.3-$m DFOT is a fast beam (f/4) optical telescope with a
pointing accuracy better than $10$ arcsec (rms). It is equipped with a
512k $\times$ 512k Andor CCD camera having a pixel size of 16 micron
and a plate scale of 0.63 arcsec per pixel. The CCD covers a field of
view of $\sim$ 5 arcmin on the sky. It is cooled thermo-electrically
to -$90$ degC and is read out at $1$ MHz speed. The corresponding
system noise is $6.1$ e- (rms) and the gain is $1.4$ e-/ADU.

The $1.2-$m OHP telescope is a f/6 optical telescope with Newtonian
focus.  The detector is a thin back-illuminated 1024 $\times$ 1024
Tektronix chip, with a read-out noise of $8.5$ e- and a gain of $3.5$
e-/ADU. This instrument provides a field of view of $\sim$
$11.7$ arcmin on the sky with a plate scale of $0.68$ arcsec per
pixel.

The exposure time for each science frame was set between $5$ and $30$
minutes, typically yielding a signal-to-noise ratio above $25-30$. The
typical seeing (FWHM) during our observations was around $2$ arcsec.
Since in the sample selection, care was taken to ensure the
availability of at least two, but usually more, comparison stars on
each CCD frame, within about $1$ mag of the target WLQ, it became
possible to identify and discount any comparison star(s) which showed
a hint of variability during a given monitoring session. Although we
  have endeavored to take comparison stars that are close to the
  monitored AGN, both in brightness and color, a deviation from this
  became necessary sometimes when trying to compare Differential Light
  Curves (DLCs) of the same AGN taken on different nights. This
  happened because of our insistence on using for a given AGN the same
  set of comparison stars on all the nights, even if different
  telescopes (having unequal fields-of-view) had been used on those
  nights. This has occasionally led to a situation when the DLCs used
  for the purpose of INOV and for `short-term optical variability
  (STOV) or `long-term optical variability (LTOV) employ different
  comparison stars for the same AGN (see,
  Table~\ref{tab_cdq_comp_inov}).

The pre-processing of the raw CCD images (bias subtraction,
flat-fielding, cosmic-ray removal and trimming) was done using the
standard tasks available in the Image Reduction and Analysis Facility
{\textsc (IRAF)} \footnote{\textsc {Image Reduction and Analysis
    Facility (http://iraf.noao.edu/). }}. The instrumental magnitudes
of the observed target AGN and the chosen comparison stars in each CCD
frame were determined by aperture
photometry~\citep{1992ASPC...25..297S, 1987PASP...99..191S}, using the
Dominion Astronomical Observatory Photometry \textrm{II} (DAOPHOT II
algorithm)\footnote{\textsc {Dominion Astrophysical Observatory
    Photometry.}}.  To select the aperture size (FWHM) for photometry,
we first determined the ``seeing" for each frame by averaging the
observed FWHMs of $5$ moderately bright stars in the CCD frame. We
then took the median of these averaged values over all the frames
recorded in the session. The aperture diameter was set equal to 2
times the median FWHM. The only exception to this procedure is the
nearby WLQ J124225.3+642919.0 ($z$ = 0.04247) for which we chose a
fixed aperture of 4 arcsec diameter in view of the fact that the
underlying fuzz due to the host galaxy is quite
prominent~\citep[see,][]{Carini1991AJ....101.1196C,Cellone2000AJ....119.1534C}. As
shown below, this object did not exhibit significant variability,
either on hour-like or longer time scale and this conclusion is not
sensitive to the size of the aperture chosen.  \par
 
To derive the DLCs of the target WLQ monitored in a given session, we
selected two steady comparison stars which are present in all the CCD
frames and are also close to the WLQ monitored, both in location and
brightness. Particulars of the comparison stars used for the various
sessions are given in Table~\ref{tab_cdq_comp_inov}. Note that with
the solitary exception of the WLQ J1532-0039, the $g-r$ color
difference between the target WLQ and its comparison stars is always
within $\sim$ 1-mag, with a median value of $0.6-mag$
(Table~\ref{tab_cdq_comp_inov}). Analysis by
~\citet{Stalin2004MNRAS.350..175S,Stalin2004JApA...25....1S} has shown
that for color difference of this order, the changing atmospheric
attenuation during a session produces a negligible effect on the DLCs
(see, also,~\citet{Carini1992AJ....104...15C}) .\par
\begin{table*}
\begin{minipage}{500mm}
{
 
  {\tiny
\caption {Basic parameters and observing dates for the $8$ WLQs and their comparison stars monitored for 
 { INOV/STOV/LTOV (sect. 5)}.
\label{tab_cdq_comp_inov}}

\begin{tabular}{rcccccccc ccc c}\\
\hline
                        & & & & &  INOV & & & STOV  & & & LTOV\\
{IAU Name} &Date{\footnote{The DLCs obtained on dates marked by $`*$' have
been used only to look for STOV/LTOV, since they are either too noisy or insufficiently\\ long (T $<$ 3 hr) for the purpose of INOV search.
The $`\dagger$' marks the dates whose WLQ DLCs using the comparison stars adopted for\\ STOV/LTOV search are not shown in Fig.~\ref{fig:lurve_inov}, as the
DLCs used for INOV search on the date are already in Fig.~\ref{fig:lurve_inov}, with the difference that\\ the comparison stars used are not the same (more optimally selected in the case of INOV search, as
described in Sect. 3)}
}    &   {R.A.(J2000)} & {Dec.(J2000)} & { g} & { r} & { g-r} & { g} & { r} & { g-r} & { g} & { r} & { g-r} \\
           &  dd.mm.yyyy    &   (h m s)       &($^\circ$ $^\prime$ $^{\prime\prime}$)   & (mag)   & (mag)   & (mag)  & (mag)   & (mag)   & (mag)  & (mag)   & (mag)   & (mag)  \\
{(1)}     &  {(2)}        & {(3)}           & {(4)}                              & {(5)}   & {(6)}   & {(7)}  & {(8)}& {(9)} & {(10)} & {(11)}& {(12)}& {(13)} \\
\hline
\multicolumn{3}{l}{}\\

J104833.5$+$620305.0&       &10 48 33.56 &$+$62 03 05.0  &              20.22 &        19.85 &        0.37\\   
S1(INOV) &02.12.2005              &10 48 05.36 &$+$62 05 19.1  &              21.31 &        19.57 &        1.74\\
S2(INOV) &02.12.2005              &10 48 35.90 &$+$61 59 29.9  &              20.63 &        19.52 &        1.11\\
S3(LTOV) &02.12.2005$^{\dagger}$,29.04.2016   &10 48 52.36 &$+$62 02 27.5  & & &  & & & & 18.52 &        17.84 &        0.68\\
S4(LTOV) &02.12.2005$^{\dagger}$,29.04.2016   &10 48 33.90 &$+$62 04 06.5  & & &  & & & & 16.74 &        16.40 &        0.34\\ \\
J121221.5$+$534128.0&       &12 12 21.56 &$+$53 41 28.0  &   19.03 &       18.63 &        0.40\\   
S1(INOV,STOV) &26.04.2006,29.04.2006   &12 12 28.97 &$+$53 43 36.4  &   19.49 & 18.48 & 1.01&   19.49 &       18.48 &        1.01\\
S2(INOV,STOV) &26.04.2006,29.04.2006   &12 11 46.56 &$+$53 40 26.9  &   18.92 &       17.94 & 0.98&   18.92 &       17.94 &        0.98\\ \\  
J123743.1$+$630144.9 &      &12 37 43.09 &$+$63 01 44.9  &   20.14 &       19.11 &        1.03\\
S1(INOV) &27.04.2006              &12 37 45.36 &$+$63 00 59.3  &   20.49 &       18.92 &        1.57\\
S2(INOV) &27.04.2006              &12 37 55.04 &$+$63 06 48.7  &   19.44 &       18.94 &        0.50\\
S3(INOV) &30.04.2006              &12 38 02.89 &$+$63 03 18.3  &   19.34 &       18.72 &        0.62\\
S4(INOV) &30.04.2006              &12 38 36.86 &$+$63 06 31.2  &   18.87 &       18.27 &        0.60\\
S5(STOV,LTOV) &27.04.2006$^{\dagger}$,30.04.2006$^{\dagger}$,11.05.2016$^{*}$ &12 38 00.29 &$+$63 02 26.4  & & &  & 17.51 & 16.87 & 0.64& 17.51 & 16.87 & 0.64\\
S6(STOV,LTOV) &27.04.2006$^{\dagger}$,30.04.2006$^{\dagger}$,11.05.2016$^{*}$  &12 37 30.63 &$+$63 02 51.5  & & &  & 18.72 & 17.98 & 0.74& 18.72 & 17.98 & 0.74\\ \\
J124225.3$+$642919.0  &     &12 42 25.39 &$+$64 29 19.0  &   16.28 &  15.69 &        0.59\\
S1(INOV,STOV) &25.04.2006,28.04.2006    &12 42 43.47 &$+$64 30 17.2  &   18.18 & 17.73 &  0.45&   18.18 &       17.73 &        0.45\\
S2(INOV,STOV) &25.04.2006,28.04.2006    &12 42 17.56 &$+$64 25 32.7  &   17.93 & 16.92 & 1.01&   17.93 &       16.92 &        1.01\\ \\
J133219.6$+$622715.9  &      &13 32 19.65 &$+$62 27 15.9  &   19.50 &       19.18 &        0.32\\
S1(INOV) &15.03.2007               &13 32 17.85 &$+$62 27 10.3  &   19.98 &       18.51 &        1.47\\
S2(INOV) &15.03.2007               &13 31 59.40 &$+$62 30 28.7  &   18.12 &       17.88 &        0.24\\
S3(INOV,STOV) &15.03.2007$^{\dagger}$,16.03.2007   &13 31 54.55 &$+$62 22 39.9  &20.57 & 19.28 & 1.29  &  20.57 &       19.28 &        1.29\\
S4(INOV,STOV) &15.03.2007$^{\dagger}$,16.03.2007   &13 32 30.17 &$+$62 28 38.0  &19.31 & 18.90 & 0.41  &  19.31 &       18.90 &        0.41\\
S1(LTOV) &16.03.2007$^{\dagger}$,05.01.2016$^{*}$   &13 32 17.85 &$+$62 27 10.3  & &&&&&&  19.98 &       18.51 &        1.47\\
S5(LTOV) &16.03.2007$^{\dagger}$,05.01.2016$^{*}$   &13 32 21.56 &$+$62 29 03.3  & & &  & & & &  18.23 &       17.70 &        0.53\\ \\
J140923.9$+$261821.1  &      &14 09 23.90 &$+$26 18 21.1  &   15.81 &       15.74 &        0.07\\
S1(INOV,LTOV) & 27.05.2016,02.05.2016$^{*}$              &14 09 31.31 &$+$26 17 49.6  &   16.27 &       15.72 &        0.55 & & & & 16.27 & 15.72 &0.55\\
S2(INOV,LTOV) & 27.05.2016, 02.05.2016$^{*}$           &14 09 59.12 &$+$26 14 26.4  &   15.73 &       15.22 &        0.51 &&& & 15.73 &       15.22 &        0.51 \\ \\
J153259.9$-$003944.1  &      &15 32 59.96 &$-$00 39 44.1  &   23.78 &       21.16 &        2.62\\
S1(INOV,STOV) & 11.04.2005,12.04.2005$^{*}$   &15 32 43.66 &$-$00 43 42.4  &19.43 & 19.12 & 0.31 &  19.43 & 19.12 &  0.31\\
   & 27.03.2017,29.03.2017$^{*}$\\
S2(INOV,STOV) & 11.04.2005,12.04.2005$^{*}$   &15 32 44.45 &$-$00 43 37.2  &18.89 & 18.79 & 0.10 &  18.89 & 18.79 &  0.10\\
   & 27.03.2017,29.03.2017$^{*}$\\ \\
J232428.4$+$144324.4  &      &23 24 28.43 &$+$14 43 24.4  &   19.04 &       18.77 &        0.27\\
S1(INOV) & 01.12.2005              &23 24 38.42 &$+$14 46 35.6  &   20.11 &       18.84 &        1.27\\
S2(INOV) & 01.12.2005              &23 24 31.08 &$+$14 47 25.4  &   20.07 &       18.64 &        1.43\\
S3(LTOV) & 01.12.2005$^{\dagger}$,10.10.2016$^{*}$ &23 24 28.79 &$+$14 42 29.7  & & &  & & & &  17.77 &       16.93 &        0.84\\
S4(LTOV) & 01.12.2005$^{\dagger}$,10.10.2016$^{*}$ &23 24 28.35 &$+$14 45 52.9  & & &  & & & &  16.99 &       16.14 &        0.85\\
\hline
\end{tabular}
}
}
\end{minipage}
\end{table*}
\section{STATISTICAL ANALYSIS}  
{ C}-statistic~\citep[e.g.,][]{1997AJ....114..565J} has been the
most commonly used and the one-way analysis of variance
(ANOVA)~\citep{Diego2010AJ....139.1269D} the most powerful test for
verifying the presence of variability in a DLC. However, we did not
employ either of these tests here
because,~\citet{Diego2010AJ....139.1269D} has questioned the validity
of the { C}-test, arguing that the { C}-statistics does not have
a Gaussian distribution and { the nominal critical value of 2.576
  used for confirming the presence of variability at $3\sigma$ level
  is usually too conservative.} On the other hand, the ANOVA test
requires a rather large number of data points in the DLC, so as to
have several points within each sub-group used for the analysis. This
is not feasible for our DLCs which typically have { only about a
  dozen data points}. So, we have instead used the \emph{F$-$test}
which is based on the ratio of variances, { F }$=
variance(observed)/variance(expected)
$~\citep{Diego2010AJ....139.1269D,Villforth2010ApJ...723..737V}.  Its
two versions are: (i) the standard { F$-$test} (hereafter
$F^{\eta}-$test,~\citet{Goyal2012A&A...544A..37G}) and (ii) scaled
$F-$test (hereafter
$F^{\kappa}-$test,~\citet{Joshi2011MNRAS.412.2717J}).  The
$F^{\kappa}-$test is preferred when the magnitude difference between
the object and comparison stars is large
~\citep{Joshi2011MNRAS.412.2717J}. Onward Paper II, we have only been
using the $F^{\eta}-$test because our choosen comparison stars are
usually quite comparable in brightness to the target AGN. An
additional gain in using the $F^{\eta}-$test is that we can directly
compare our INOV results with those now known for other major AGN
classes ~\citep{Goyal2013MNRAS.435.1300G}. An important point to keep
in mind while applying the statistical tests is that the photometric
errors on individual data points in a given DLC, as returned by the
routine in the IRAF and DAOPHOT software are systematically too low by
a factor $\eta$ ranging between $1.3$ and $1.75$, as estimated in
independent studies~\citep[e.g.,][] {1995MNRAS.274..701G,
  1999MNRAS.309..803G, Sagar2004MNRAS.348..176S,
  Stalin2004JApA...25....1S, Bachev2005MNRAS.358..774B}. Recently,
using a large sample, ~\citet{Goyal2013MNRAS.435.1300G} estimated the
best-fit value of $\eta$ to be $1.5$, which is adopted here. Thus, the
$F^{\eta}-$ statistics can be expressed as:
\begin{equation} 
 \label{eq.fetest}
F_{1}^{\eta} = \frac{\sigma^{2}_{(q-s1)}}
{ \eta^2 \langle \sigma_{q-s1}^2 \rangle}, \nonumber  \\
\hspace{0.2cm} F_{2}^{\eta} = \frac{\sigma^{2}_{(q-s2)}}
{ \eta^2 \langle \sigma_{q-s2}^2 \rangle},\nonumber  \\
\end{equation}
where $\sigma^{2}_{(q-s1)}$ and $\sigma^{2}_{(q-s2)}$ are the
variances of the `quasar-star1' and `quasar-star2' DLCs, { with
  $\langle \sigma_{q-s1}^2
  \rangle=\sum_\mathbf{i=0}^{N}\sigma^2_{i,err}(q-s1)/N$ and $\langle
  \sigma_{q-s2}^2
  \rangle=\sum_\mathbf{i=0}^{N}\sigma^2_{i,err}(q-s2)/N$} being the
mean square (formal) rms errors of the individual data points in the
`quasar-star1' and `quasar-star2' DLCs, respectively. Note that the
number of points (N) in all the DLCs for a given session is
practically the same and the scaling factor $\eta$ is taken to be
=$1.5$, as mentioned above.

The $F^{\eta}$-test is applied to a given `quasar-star' DLC by
computing its $F$ value using Eq.~\ref{eq.fetest}, and then comparing
it with the critical value, $F^{(\alpha)}_{\nu_{qs}}$, where $\alpha$
is the significance level set for the test, and $\nu_{qs}$ is the
degree of freedom (N $-$ 1) of the DLC.  The two values we have set
for the significance level are $\alpha=$ 0.01 and 0.05, which
correspond to confidence levels of greater than 99 and 95 per cent,
respectively. If the computed $F$ value exceeds the corresponding
critical value $F_{c}$, the null hypothesis (i.e., no variability) is
discarded to the respective level of confidence. Thus, we denote a
science target as \emph{variable} (`V') in a given session, if the
computed $F$-values for both its DLCs are $\ge F_{c}(0.99)$, which
corresponds to a confidence level $\ge 99$ per cent, and term it as
\emph{non- variable} (`NV') if either of the two DLCs is found to have
an $F$-value $\le F_{c}(0.95)$. The remaining cases are classified as
\emph{probably variable} (`PV').

The inferred INOV status of the DLCs of each WLQ, relative to its two
chosen comparison stars, are presented in Table~\ref{wl:tab_res} for
each monitoring session. In the first 4 columns, we list the name of
the WLQ, the date and duration of its monitoring and the number of
data points (N) (which is the same for both the DLCs of the WLQ). The
next two columns list the computed $F$-values for the two
`quasar$-$star' DLCs and their INOV status, based on the
$F^{\eta}-$test. Column $7$ gives our averaged photometric error
$\sigma_{i,err}(q-s)$ of the data points in the two `quasar$-$star'
DLCs.  Typically, it is 0.03-mag for these relatively faint objects.
\begin{table*}
\centering
\begin{minipage}{500mm}
{\small
\caption{Observational details and the inferred INOV status for the present sample of 8 WLQs monitored 
in 13 sessions.}
\label{wl:tab_res}
\begin{tabular}{@{}ccc c rrr rrr cc@{}}
\hline \multicolumn{1}{c}{RQWLQ} &{Date} &{T} &{N} 
&\multicolumn{1}{c}{F-test values} 
&\multicolumn{1}{c}{INOV status{\footnote{V=variable, i.e., confidence level
$\ge 0.99$ for both DLCs; PV=probable variable, i.e., $0.95-0.99$ confidence level for both DLCs;
\\NV=non-variable, i.e., confidence level $< 0.95$ for one or both DLCs.
Variability status inferred using the\\ quasar-star1 and quasar-star2
pairs are separated by a comma.}}}
&{$\sqrt { \langle \sigma^2_{i,err} \rangle}$}\\
& dd.mm.yyyy& 
hr & &{$F_1^{\eta}$},{$F_2^{\eta}$}
&F$_{\eta}$-test &(q-s)\\
(1)&(2) &(3) &(4) &(5)&(6)
&(7)\\ \hline
J104833.5$+$620305.0 & 02.12.2005& 4.95& 17& 1.12,1.42 & NV,NV&0.04\\
J121221.5$+$534128.0 & 26.04.2006& 5.78& 12& 0.71,0.64 & NV,NV&0.03\\
J121221.5$+$534128.0 & 29.04.2006& 5.28& 11& 0.24,0.34 & NV,NV&0.04\\
J123743.1$+$630144.9 & 27.04.2006& 4.20&  9& 0.30,0.16 & NV,NV&0.03\\
J123743.1$+$630144.9 & 30.04.2006& 5.73& 12& 0.76,0.63 & NV,NV&0.05\\
J124225.3$+$642919.0 & 25.04.2006& 4.65& 18& 0.64,1.56 & NV,NV&0.01\\
J124225.3$+$642919.0 & 28.04.2006& 3.29& 13& 0.88,1.25 & NV,NV&0.01\\
J133219.6$+$622715.9 & 15.03.2007& 4.79& 10& 5.24,4.84 & PV,PV&0.03\\
J133219.6$+$622715.9 & 16.03.2007& 6.90& 13& 0.59,0.49 & NV,NV&0.02\\
J140923.9$+$261821.1 & 27.05.2016& 3.66& 40& 0.37,0.54 & NV,NV&0.01\\
J153259.9$-$003944.1 & 11.04.2005& 4.50& 15& 0.44,0.48 & NV,NV&0.08\\
J153259.9$-$003944.1 & 27.03.2017& 4.65& 10& 1.49,1.59 & NV,NV&0.08\\
J232428.4$+$144324.4 & 01.12.2005& 2.92& 10& 0.81,1.04 & NV,NV&0.02\\
\hline
\end{tabular}
}
\end{minipage}
\end{table*} 
 \begin{figure*}
 \centering
 \psfig{figure=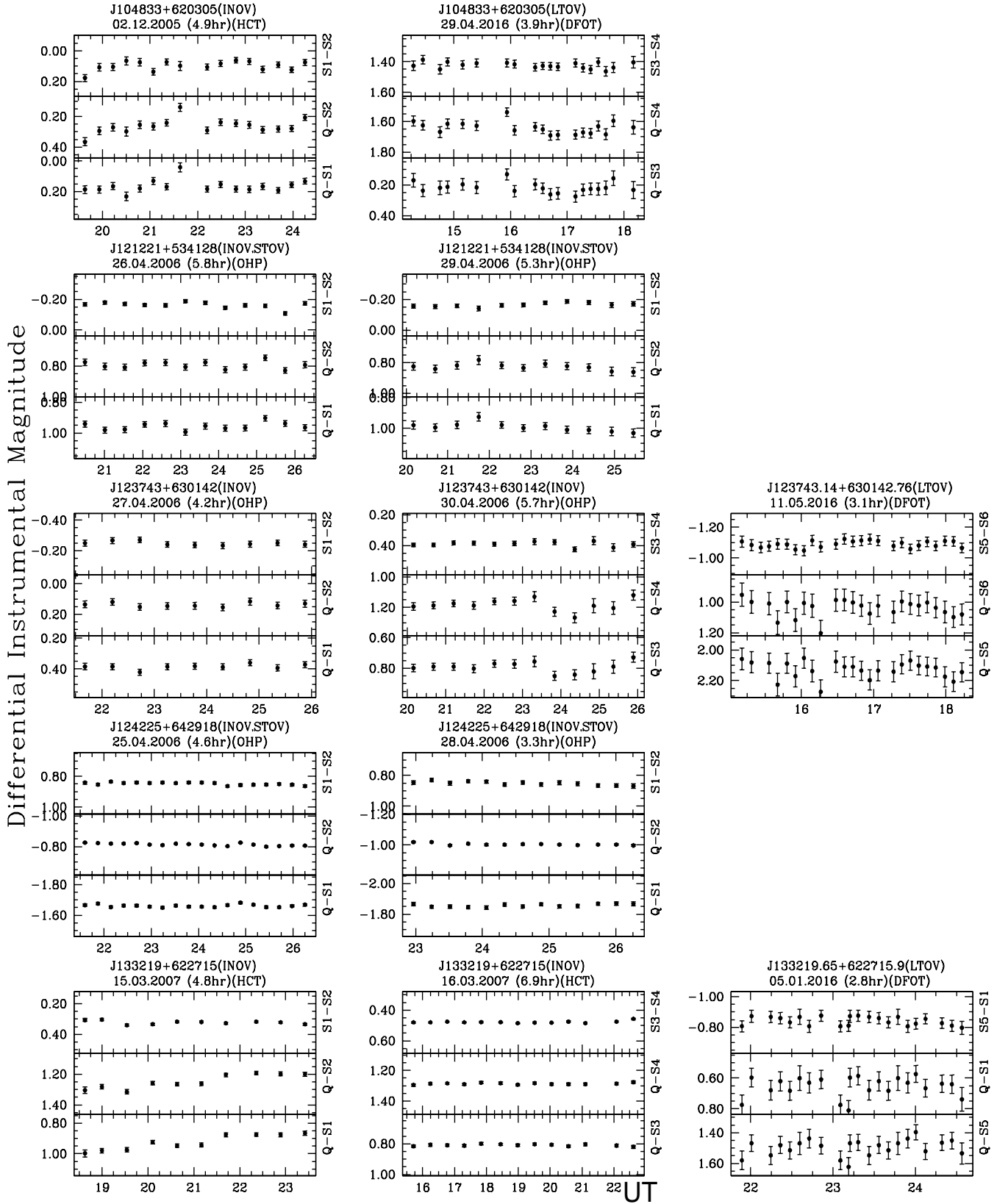,height=21.8cm,width=20.0cm,angle=00,bbllx=20bp,bblly=142bp,bburx=580bp,bbury=715bp,clip=true}
 \vspace{-0.5cm}
 \caption[]{Differential light curves (DLCs) in R-band, for the $8$
   WLQs in our sample.  Name of the WLQ along with the date and
   duration of its monitoring session,  are given at the top of each 
   panel together with the { telescope's name and the type of variability for which the DLCs were used}. 
   Within each panel the uppermost DLC is
   derived using the light curves of the two comparison stars, while
   the lower two DLCs are the light curves of the WLQ relative to the
   two comparison stars, as defined in the labels just outside the
   panel on the right side. \emph{(cont.)}}
 \label{fig:lurve_inov}
  \end{figure*}
\begin{figure*} 
 \centering
 \psfig{figure=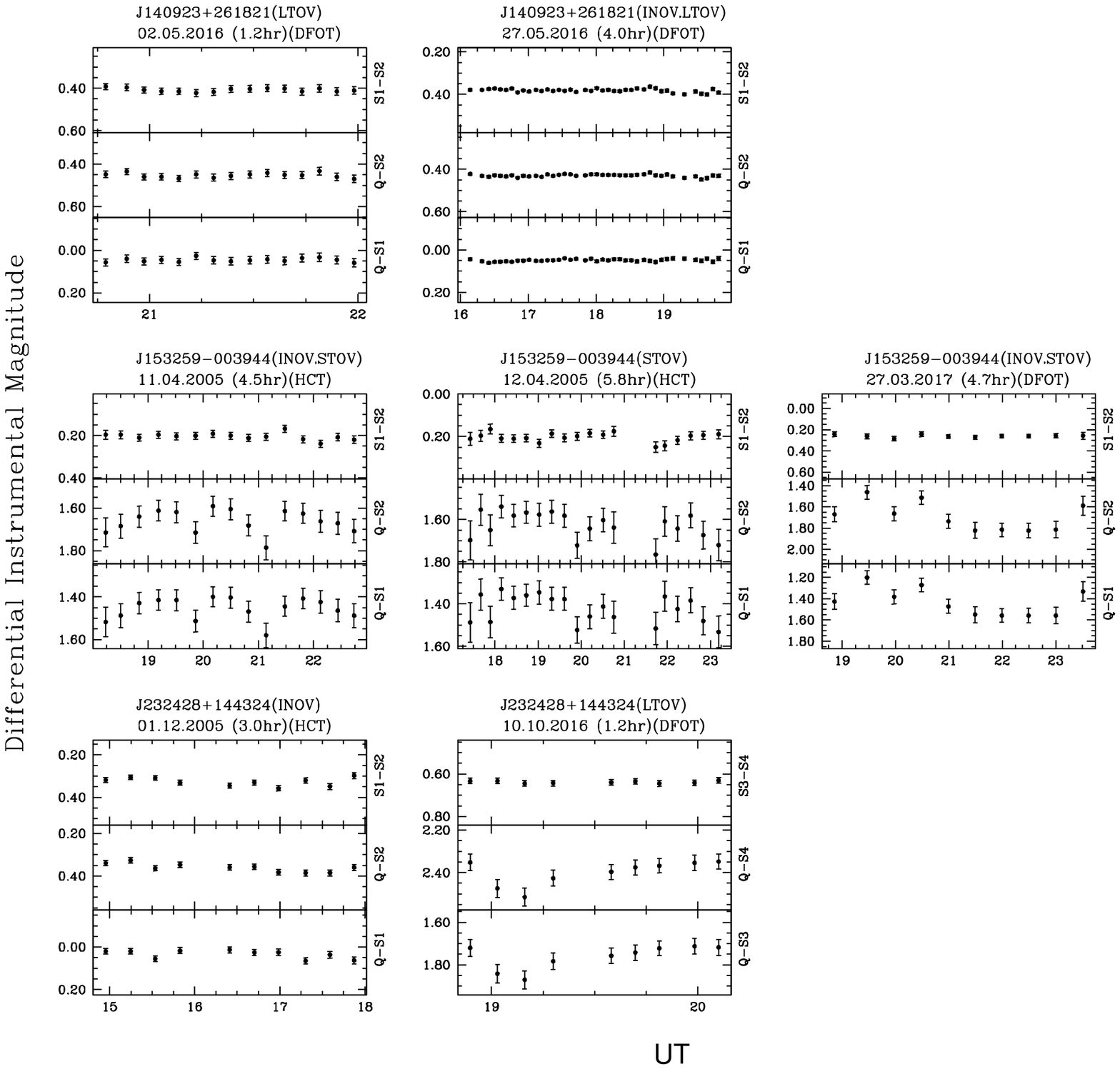,height=16.5cm,width=17.0cm,angle=00,bbllx=20bp,bblly=142bp,bburx=580bp,bbury=715bp,clip=true} 
\vspace{-0.8cm}
 \contcaption{}
 \end{figure*}
\begin{table*}
\begin{minipage}{500mm}
{
\caption{Session-averaged brightness (in unit of instrumental magnitude) of the DLCs of the $8$ WLQs and their comparison stars on two epochs.
\label{tab:Var_info}}
\begin{tabular}{crrrc} 
\hline
\multicolumn{1}{c}{Source Name} &Averaged brightness(epoch 1,Telescope)& Averaged brightness(epoch 2,Telescope) & Brightness change (mag)\\
(variability type) \\
\hline

J104833.5$+$620305.0\\
Q-S3(LTOV)  &$1.892\pm0.008$(02.12.2005,HCT) &$0.217\pm0.008$ (29.04.2016,DFOT) &$-1.675\pm0.011$\\
Q-S4(LTOV)  &$3.275\pm0.008$(02.12.2005,HCT) &$1.641\pm0.009$ (29.04.2016,DFOT) &$-1.634\pm0.012$\\
S3-S4(LTOV)  &$1.382\pm0.001$(02.12.2005,HCT) &$1.424\pm0.004$ (29.04.2016,DFOT) &$0.042\pm0.004$\\
J121221.5$+$534128.0\\
Q-S1(STOV)  &$0.955\pm0.007$(26.04.2006,OHP) &$0.992\pm0.007$(29.04.2006,OHP) &$0.037\pm0.009$\\
Q-S2(STOV)  &$0.793\pm0.006$(26.04.2006,OHP) &$0.826\pm0.005$(29.04.2006,OHP) &$0.033\pm0.007$\\
S1-S2(STOV) &$-0.162\pm0.005$(26.04.2006,OHP) &$-0.166\pm0.003$(29.04.2006,OHP) &$-0.004\pm0.005$\\
J123743.1$+$630144.9\\
Q-S5(STOV)& $2.305\pm0.004$(27.04.2006,OHP)&$2.300\pm0.008$(30.04.2006,OHP)&$-0.005\pm0.008$\\
Q-S6(STOV)& $1.186\pm0.004$(27.04.2006,OHP)&$1.186\pm0.002$(30.04.2006,OHP)&$0.000\pm0.012$\\
S5-S6(STOV)& $-1.118\pm0.003$(27.04.2006,OHP)&$-1.110\pm0.006$(30.04.2006,OHP)&$0.008\pm0.006$\\
\\
Q-S5(LTOV)& $2.300\pm0.008$(27.04.2006,OHP)&$2.126\pm0.010$(11.05.2016,DFOT)&$-0.174\pm0.012$\\
Q-S6(LTOV)& $1.186\pm0.012$(27.04.2006,OHP)&$1.035\pm0.010$(11.05.2016,DFOT)&$-0.151\pm0.015$\\
S5-S6(LTOV)& $-1.110\pm0.006$(27.04.2006,OHP)&$-1.090\pm0.004$(11.05.2016,DFOT)&$0.020\pm0.007$\\
J124225.3$+$642919.0\\
Q-S1(STOV)& $-1.887\pm0.002$(25.04.2006,OHP)&$-1.885\pm0.002$(28.04.2006,OHP)&$0.002\pm0.002$\\
Q-S2(STOV)& $-1.035\pm0.002$(25.04.2006,OHP)&$-1.034\pm0.001$(28.04.2006,OHP)&$0.001\pm0.002$\\
S1-S2(STOV)& $0.851\pm0.003$(25.04.2006,OHP)&$0.851\pm0.003$(28.04.2006,OHP)&$0.000\pm0.004$\\
J133219.6$+$622715.9\\
Q-S3(STOV)& $0.767\pm0.012$(15.03.2007,HCT)&$0.808\pm0.001$(16.03.2007,HCT)&$0.041\pm0.012$\\
Q-S4(STOV)& $1.241\pm0.013$(15.03.2007,HCT)&$1.286\pm0.001$(16.03.2007,HCT)&$0.045\pm0.013$\\
S3-S4(STOV)& $0.473\pm0.003$(15.03.2007,HCT)&$0.478\pm0.001$(16.03.2007,HCT)&$0.005\pm0.003$\\
\\
Q-S5(LTOV)& $1.466\pm0.001$(16.03.2007,HCT)&$1.492\pm0.009$(05.01.2016,DFOT)&$0.026\pm0.009$\\
Q-S1(LTOV)& $0.808\pm0.001$(16.03.2007,HCT)&$0.655\pm0.011$(05.01.2016,DFOT)&$-0.153\pm0.011$\\
S5-S1(LTOV)& $-0.658\pm0.001$(16.03.2007,HCT)&$-0.836\pm0.005$(05.01.2016,DFOT)&$-0.178\pm0.005$\\
J140923.9$+$261821.1\\
Q-S1(LTOV)& $0.045\pm0.002$(02.05.2016,DFOT)&$0.048\pm0.001$(27.05.2016,DFOT)&$0.003\pm0.002$\\
Q-S2(LTOV)& $0.453\pm0.002$(02.05.2016,DFOT)&$0.429\pm0.001$(27.05.2016,DFOT)&$-0.024\pm0.002$\\
S1-S2(LTOV)& $0.407\pm0.002$(02.05.2016,DFOT)&$0.381\pm0.001$(27.05.2016,DFOT)&$-0.026\pm0.002$\\
J153259.9$-$003944.1\\
Q-S1(STOV)& $1.450\pm0.012$(11.04.2005,HCT)&$1.411\pm0.014$(12.04.2005,HCT)&$-0.039\pm0.018$\\
Q-S2(STOV)& $1.654\pm0.013$(11.04.2005,HCT)  &$1.615\pm0.013$(12.04.2005,HCT) &$-0.039\pm0.018$\\
S1-S2(STOV)& $0.204\pm0.004$(11.04.2005,HCT) &$0.203\pm0.004$(12.04.2005,HCT)&$-0.001\pm0.005$\\
\\
Q-S1(STOV)& $1.430\pm0.041$(27.03.2017,DFOT)&$1.661\pm0.112$(29.03.2017,DFOT)&$0.231\pm0.119$\\
Q-S2(STOV)& $1.690\pm0.042$(27.03.2017,DFOT)&$1.906\pm0.111$(29.03.2017,DFOT) &$0.216\pm0.118$\\
S1-S2(STOV)& $0.260\pm0.003$(27.03.2017,DFOT)&$0.244\pm0.028$(29.03.2017,DFOT)&$-0.016\pm0.028$\\
J232428.4$+$144324.4\\
Q-S3(LTOV)& $1.624\pm0.003$(01.12.2005,HCT)&$1.757\pm0.017$(10.10.2016,DFOT)&$0.133\pm0.017$\\
Q-S4(LTOV)& $2.240\pm0.002$(01.12.2005,HCT)&$2.394\pm0.018$(10.10.2016,DFOT) &$0.154\pm0.018$\\
S3-S4(LTOV)& $0.616\pm0.001$(01.12.2005,HCT)&$0.637\pm0.001$(10.10.2016,DFOT) &$0.021\pm0.001$\\
\hline
\end{tabular}
}
\end{minipage}
\end{table*}
\section{Results}

As seen from Table~\ref{tab_cdq_comp_inov}, we have monitored the $8$
WLQs in the present sample on $20$ nights. Of these, 7 nights' data
could only be used to look for day-like, or long-term variability,
since either the individual data points are too noisy (4 nights), or
the DLCs are of insufficient duration (T $<$ 3 hr) for the purpose of
INOV search.

\subsection {Remarks on possible INOV events}
{\bf DLCs} derived for the $19$ intra-night monitoring sessions
covering the 8 WLQs are displayed in Fig.~\ref{fig:lurve_inov}, except
the brief session of J153259.9$-$003944.1 on 29-03-2017 with DFOT
which gave just two points and which is hence used only for STOV
search. The top of each panel provides information on the telescope
used, duration of the DLCs, and the type of variability searched using
those DLCs.  Among the 13 sessions used for INOV search, the telescope
HCT was used for 5 of the sessions, OHP for 6 sessions and DFOT for 2
session. In each panel, the upper plot shows the star-star DLC for
that session, while the lower two plots are DLCs of the WLQ relative
to the two (comparison) stars. Note that, except for
J153259.9$-$003944.1, the faintest object in the present sample
(Table~\ref{tab:source_info}), at least one of the two comparison
stars is within a magnitude of the WLQ monitored. The rms noise on
individual data points in the DLCs varies from session to session and
even within a session, depending on the changing observing conditions,
such as the varying sky brightness due to the Moon, etc. Nonetheless,
barring the case of WLQ J153259.9$-$003944.1, the typical rms noise of
individual data points is 0.03 mag (Table~\ref{wl:tab_res}). Although,
based on the statistical test mentioned in Sect. 4, there is no formal
detection of INOV during any of the 13 sessions
(Table~\ref{wl:tab_res}), a visual inspection of the DLCs in
Fig.~\ref{fig:lurve_inov} provides a reasonable hint of INOV in the
following two sessions.\\ \textbf{J104833.56$+$620305.0 (02-12-2005):}
For the first half an hour, the dip seen in the upper two panels (q-s2
and s1-s2 DLCs) is clearly due to variation of the star S2.  The most
prominent feature of the DLCs is an $\approx$ 0.1-mag spike at $\sim$
21.6 UT, in both q-s1 and q-s2 DLCs, while the s1$-$s2 DLC is steady
over that time span. Here it is pertinent to note that although a
photometric redshift ($z \sim 1.54$) has been reported for this
source~\citep{Richards2009ApJS..180...67R} its extragalactic nature
remains uncertain. This is, firstly, in view of a hint of proper
motion (Table~\ref{tab:source_info}), but more so because of the lack
of any features in its optical spectrum ~\citep[e.g.,
  see][]{Plotkin2010AJ....139..390P,Kuegler2014A&A...569A..95K}.  For
optical polarization of this source,~\citet{Smith2007ApJ...663..118S}
have given an upper limit of $2.33\%$. Clearly, a proper
interpretation of the INOV results for this source remains contingent
upon the availability of a secure measurements of its redshift and/or
proper motion. Furthermore, the radio-optical spectral index
$\alpha_{r-o}$ ~\citep[$<$ 0.27,][] {Collinge2005AJ....129.2542C} for
this source falls a bit short of the usually accepted upper limit of
0.21 for radio-quiet
quasars~\citep[e.g.,][]{Luo2015ApJ...805..122L}. Therefore, in order
to be certain of its radio-quiet classification, radio flux
measurement is needed with a detection threshold a few times lower
than attained in the FIRST survey~\citep{Becker1995ApJ...450..559B},
which is currently the deepest radio measurement available for this
source.\\ \textbf{J133219.65+622715.9 (15-03-2007):} For this
high-redshift WLQ ($z = 3.1514$, Table~\ref{tab:source_info})
~\citet{Collinge2005AJ....129.2542C} give $\alpha_{r-o}$ = 0.19, which
places it near the borderline between radio-quiet and radio-loud
quasars. Indeed, with a radio-louness parameter R =
33~\citep{Lane2011ApJ...743..163L}, this distant WLQ qualifies to be a
"Radio Intermediate Quasar
(RIQ)"~\citep[e.g.,][]{Miller1993MNRAS.263..425M,Falcke1995A&A...298..375F,
  Falcke1996ApJ...473L..13F,Barvainis2005ApJ...618..108B}.  A well
known local prototype of this class is III Zw 2 which is known to
exhibit a blazar-like behaviour, i.e., a large radio
variability~\citep{Teraesranta1998A&AS..132..305T}, a superluminal
motion~\citep{Brunthaler2000A&A...357L..45B,
  Brunthaler2005A&A...435..497B}, as well as $\gamma-$ray flaring in
the radio core~\citep{Liao2016ApJS..226...17L}. In fact, based on
rapid radio flux variability, relativistically beamed jets are
inferred to be common among
RIQs~\citep{Wang2006ApJ...645..856W}. However, a study of the optical
variability of RIQs on hour to day-like time scales, based on
intranight monitoring of 10 RIQs on 42 nights, has shown that while
intra-night and inter-night optical variability is exhibited by RIQs,
the INOV amplitude hardly ever exceeds $3-4\%$, in clear contrast to
blazars~\citep{Goyal2010MNRAS.401.2622G}. This is consistent with the
inference that a relativistically beamed radio core of a quasar is not
a sufficient condition for observing a pronounced
INOV~\citep{Goyal2012A&A...544A..37G}.\\ Broadly, the trend visually
inferred from each DLC of this WLQ is a $\sim$ 0.1-mag brightening
over the 5 hour session on 15.03.2007, relative to both comparison
stars which themselves remained steady (Fig.~\ref{fig:lurve_inov};
Table~\ref{wl:tab_res}). Additional information on rapid optical
variability of this source is provided in Sect. 6.1. It should be
mentioned that the source was found to be unpolarised (p $<$ 1.7\%) in
the optical, which corresponds to UV in the rest-frame
~\citep{Heidt2011A&A...529A.162H}.\\
\subsection {Notes on inter-day variability events}
For 5 WLQs in the present sample, the observations reported here also
enable us to investigate STOV on day-like time scale
(Table~\ref{tab_cdq_comp_inov}). Each of these sources was monitored
on two epochs separated by 1 to 3 days. We emphasize that for
computing the STOV (or LTOV) status of a WLQ, as reported in
Table~\ref{tab:Var_info}, we have used the same pair of comparison
stars for all the monitoring sessions. In each case, the stars were
found to remain steady, but the session-averaged brightness showed a
small, yet significant change which is 0.035$\pm$0.008 mag for
J121221.5$+$534128.0; 0.043$\pm$0.009 for J133219.6$+$622715.9 and
0.039$\pm$0.013 $\&$ 0.223$\pm$0.083 for J153259.9$-$003944.1 in the
two STOV sessions (Table~\ref{tab:Var_info}). { Applying the F-test on
  the combined DLCs, we found that only J133219$+$622715 shows
  variability at more than $99\%$ confidence level.}  { This is
  further discussed in Sect. 6.}
\subsection {Variability on month/year-like time scales}
For 5 of our WLQs, the observations reported here permit investigation
of LTOV (see Table~\ref{tab_cdq_comp_inov}). As seen from {
  Table~\ref{tab:Var_info}}, LTOV is clearly witnessed for
J104833.5$+$620305.0, J123743.1$+$630144.9 and J232428.4$+$144324.4,
relative to both their comparison stars. Unfortunately, the large
variability of 1.655$\pm$0.009 mag observed for J104833.5$+$620305.0
is difficult to interpret at present, since extragalactic nature of
this source remains to be established (Sect. 5.1). Thus, in the
present sample only two sources viz, J123743.1$+$630144.9 ($z$ =
3.4278) and J232428.4$+$144324.4 ($z$ = 1.4172) showing level changes
over a decade of 0.163$\pm$0.010 mag and 0.144$\pm$0.018 mag,
respectively, are at present clear examples of a radio-quiet WLQ
displaying a substantial long-term variability. A previously known
such example is J153259.9$-$003944.1 ($z$ = 4.62) for which $i-$ band
photometry spanning 7-years has revealed a variability of amplitude
$\sim$0.1-mag~\citep{DiamondStanic2009ApJ...699..782D}. Note that for
the RQWLQs J153259.9$-$003944.1 and J123743.1$+$630144.9,
~\citet{Lane2011ApJ...743..163L} have presented SEDs covering the
rest-frame mid-IR to UV, and have inferred them to be fully consistent
with radio-quiet quasars and displaying no features characteristic of
BL Lacs.
\section{Discussion and Conclusions}
From early on it has been recognized that a multi-pronged probe is the
key to unraveling the nature and composition of the WLQ population
which is small but continues to be
enigmatic~\citep[e.g.,][]{Wu2011ApJ...736...28W}.  The strategies
employed include measurement of multi-wavelength spectral, as well as
polarimetric and temporal properties. Our investigations are focused
on the last tool, namely the temporal properties of the broad-band
optical continuum, going down to hour-like time scale. The present
study reports intra-night differential light curves (DLCs) for $8$
RQWLQs monitored in $20$ sessions, longer than $3$ hours (barring the
three sessions on 02-05-2016, 10-10-2016 and 29-03-2017 devoted to the
RQWLQs J140923.9$+$261821.1, J232428.4$+$144324.4 and
J153259.9$-$003944.1, respectively, which lasted for just $\sim$1.5
hour (or even less) and hence used for STOV/LTOV search only, (see
Fig.~\ref{fig:lurve_inov} and Table~\ref{tab_cdq_comp_inov}). Out of
the $8$ radio-quiet/weak WLQs, six have been classified as
`radio-quiet probable BL Lacs' by~\citet{Collinge2005AJ....129.2542C},
while the remaining two are reported to have exhibited an episode of
large, blazar-like optical variability (Sect. 2). These two RQWLQs
are: J140923.9+261821.1~\citep{Blundell2003ApJ...591L.103B} and
J153259.9$-$003944.1 ~\citep{Stalin2005MNRAS.359.1022S}. As in Papers
I - IV, an INOV detection threshold of $\sim 0.1-$mag was typically
achieved for $13$ of our intra-night sessions of $>$ 3 hours, covering
these $8$ sources (Table~\ref{wl:tab_res}). As mentioned in Sect. 5,
even though, based on formal statistical test, none of the $13$
sessions exhibited INOV, a visual inspection does provide a hint of
INOV in two of the sessions, namely on 02-12-2005
(J104833.5$+$620305.0) and on 15-03-2007
(J133219.6$+$622715.9). Interestingly, the J133219.6$+$622715.9
is a high redshift `radio-intermediate' WLQ and the same
classification cannot at present be excluded even for the
  J104833.5$+$620305.0, given that the existing upper limit to its
radio flux falls a few times short of that needed to confirm radio
quietness (Sect. 5.1). Furthermore, the extragalactic
classification of J104833.5$+$620305.0 remains unestablished,
since no feature has been detected in its optical spectra, thus
underscoring the need for a deeper spectroscopy (Sect. 5.1). The low
rate of INOV detection for the present sample of WLQs is in conformity
with our previously published INOV results for an independent, larger
sample of RQWLQs for which just two incidences of INOV were found
(Papers I to IV; Sect. 1).

 For five members of our sample of radio-quiet/weak WLQs, the present
 multi-epoch intra-night monitoring has also enabled a sensitive
 search for STOV on day-like time scale, and LTOV on
 month/year-like time scale. The present study of STOV/LTOV has the
 merit that for a given WLQ, the DLCs of the intranight sessions being
 compared were derived using the same comparison stars. Also, for each
 WLQ, either both comparison stars were found to have the same
 brightness in the two sessions, or at least one of the stars could be
 clearly identified as the steady one. The results are summarized
 below.

\subsection{Short-term Optical Variability on day-like time scale}
{ For the 3 of the 5 WLQs, the observed difference between the
  session-averaged brightness levels show a statistically significant
  STOV, the variation on day-like time scale being 0.035$\pm$0.008 mag for
  J121221.5$+$534128.0; 0.043$\pm$0.009 mag for J133219.6$+$622715.9;
  and 0.039$\pm$0.013 mag $\&$ 0.223$\pm$0.083 mag for
  J153259.9$-$003944.1 ({ Table~\ref{tab:Var_info}}). In the
  rest-frame, the time scale for these 4 STOV events lie between 5 and
  18 hours. While, as mentioned above, J133219.6$+$622715.9 is a
  radio-intermediate type, the other two WLQs are bona-fide RQWLQs,
  (Table~\ref{tab:source_info}). It is instructive to compare the
  inter-day variability rate of these 3 WLQs with the ensemble
  variability of optically selected quasars (i.e., overwhlemingly
  RQQs) on day-like time scale. Such a `control sample' of typical
  quasars can, for instance, be built using the analysis of the
  $R$-band monitoring of ~28,000 luminous, broad-line AGNs from the
  SDSS, under the `intermediate Palomar Transient Factory' (iPTF) and
  the `Palomar Transient Factory' (PTF)
  surveys~\citep{Caplar2017ApJ...834..111C}. Using this huge dataset,
  these authors have shown that the amplitude of variability exhibits
  a clear { anti-correlation} with luminosity and little or no
  variation with black-hole mass or redshift. Further, for the
  (radio-quiet) QSOs in the highest redshift bin ($< $z$ >$ = 2.17) of
  their sample (which approach fairly closely our 3 WLQs in
  luminosity), they find a variability amplitude of at most $1-2\%$ on
  1-day time scale in the rest-frame (section 4.1 of their
  paper). This is consistent with the rest-frame value of $\sim 1\%$
  estimated for intra-day optical variability of radio-quiet QSOs in
  the LINEAR survey~\citep{Ruan2012ApJ...760...51R}. These variability
  amplitudes for typical radio-quiet QSOs are somewhat smaller,
  although probably not inconsistent with the values given above for
  the 3 WLQs, namely J121221.5$+$534128.0 ($z$ = 3.0976),
  J133219.6$+$622715.9 ($z$ = 3.1514) and J153259.9$-$003944.1 ($z$ =
  4.62). Also, note that for blazars, ~\citet{Ruan2012ApJ...760...51R}
  give a typical intra-night amplitude of $\sim 5\%$ (see above), very
  similar to the typical value evident from the INOV amplitudes
  reported by~\citet{Goyal2013MNRAS.435.1300G} for a large sample of
  24 blazars monitored in $85$ sessions of around 6 hour
  duration each. Still, in our view, it would be pre-mature to associated
  the observed (rest-frame) intra-day variability of these 3 high-$z$
  WLQs with the blazar phenomenon particularly because the variability is
  intrinsically in the far-UV band.  This is relevant since normal
  QSOs are found to be distinctly more variable in the rest-frame UV,
  and even more in the far-UV~\citep{Welsh2011A&A...527A..15W,
    Edelson1996ApJ...470..364E, Paltani1998A&A...340...47P,
    Collier2001ApJ...561..146C}. Conceivably, the larger UV variations
  are driven by X-ray flares above the inner accretion disk, which are
  processed inside the disk~\citep[e.g.,][and references
    therein]{Rokaki1993A&A...272....8R,Wiita1996ASPC..110...42W,
    Liu2016MNRAS.462L..56L}.}\par

Finally, turning to longer time scales, the present observations have
enabled a comparison of optical brightness at rest-frame time
intervals between 2 to 5 years for four of our WLQs and of $\sim$ 2
weeks for J140923.9$+$261821.1 (Table~\ref{tab:Var_info}). Down to
2-3$\%$ limit, no long-term variability was detected for the
radio-intermediate WLQ J133219.6$+$622715.9, and the well known
radio-quiet WLQ J140923.9$+$261821.1 (Sect. 2;
Table~\ref{tab:source_info}). In contrast, the other 3 WLQs, namely
J104833.5$+$620305.0, J123743.1$+$630144.9 and J232428.4+144324.4 were
found to vary prominently over a rest-frame time interval of $\sim$ 4,
2 and 4 yr respectively (Table~\ref{tab:Var_info}). In particular, the
variability of 1.655$\pm$0.009 mag exhibited by J104833.5+620305.0
appears blazar-like, since on such time scales, optical variability is
mostly limited to $\sim$ 0.4-mag even for BL Lacs belonging to
optically selected samples; variability amplitudes above 1.5-mag are
extremely rare ~\citep[e.g.,][]{Kugler2014A&A...569A..95K}. This makes
it all the more important to ascertain the extragalactic as well as
radio-quiet classification for this object through more sensitive
optical spectroscopy and deeper radio measurements reaching a
detection threshold of $\sim 0.1$ mJy (Sect. 5). As of now, the only
two confirmed radio-quiet WLQ in our sample, for which a large
long-term variability (0.163$\pm$0.010 mag and 0.144$\pm$0.018 mag,
over a decade) is claimed here are J123743.1+630144.9 ($z=$3.4278) and
J232428.4+144324.4 ($z=$1.1472), respectively. It is also pertinent to
note that optical polarimetry of all these three sources, viz
J104833.5$+$620305.0, J123743.1$+$630144.9 and J232428.4+144324.4, has
yielded a low degree (p $<$ $\sim$ 2$\%$) of polarization
~\citep{Smith2007ApJ...663..118S,Heidt2011A&A...529A.162H}, which
appears too modest to justify a BL Lac classification. However, strong
polarization variability being the hallmark of BL Lacs, even bona-fide
BL Lac objects may sometimes appear weakly polarised ( $p$ $<$
3\%)~\citep[e.g.,][]{Wills1992ApJ...398..454W}. According
to~\citet{Fugmann1988A&A...205...86F}, there is $\sim$ 40\% chance
that a bona-fide blazar will exhibit $p$ $<$ 3\% at a random
epoch. Likewise, in another, independent study it was shown that the
fraction of time an X-ray selected BL Lac exhibits $p$ $>$ 4\% is only
$\sim$ 50\% ~\citep{Jannuzi1994ApJ...428..130J}. Therefore, the
single-epoch measurements of $p$ $<\sim$ 2\% do not rule out a BL Lac
classification, { per se} (note that even the very prominent BL Lac
object OJ 287 has been found unpolarized sometimes, e.g., see
~\citet{Villforth2009MNRAS.397.1893V}). A multi-epoch optical
polarimetry will clearly be needed to ascertain a BL Lac
classification for these interesting objects.
\section{Acknowledgements}

We are very thankful to an anonymous referee for insightful comments
on the manuscript. G-K is partly supported by a Platinum Jubilee
Senior Scientist Fellowship of the National Academy of Sciences,
India.

\bibliography{references}

\begin{thebibliography}{112}
\expandafter\ifx\csname natexlab\endcsname\relax\def\natexlab#1{#1}\fi

\bibitem[{{Abazajian} {et~al}\mbox{.}(2004){Abazajian}, {Adelman-McCarthy},
  {Ag{\"u}eros}, {Allam}, {Anderson}, {Anderson}, {Annis}, {Bahcall}, {Baldry},
  {Bastian}, {Berlind}, {Bernardi}, {Blanton}, {Bochanski}, {Boroski},
  {Briggs}, {Brinkmann}, {Brunner}, {Budav{\'a}ri}, {Carey}, {Carliles},
  {Castander}, {Connolly}, {Csabai}, {Doi}, {Dong}, {Eisenstein}, {Evans},
  {Fan}, {Finkbeiner}, {Friedman}, {Frieman}, {Fukugita}, {Gal}, {Gillespie},
  {Glazebrook}, {Gray}, {Grebel}, {Gunn}, {Gurbani}, {Hall}, {Hamabe},
  {Harris}, {Harris}, {Harvanek}, {Heckman}, {Hendry}, {Hennessy}, {Hindsley},
  {Hogan}, {Hogg}, {Holmgren}, {Ichikawa}, {Ichikawa}, {Ivezi{\'c}}, {Jester},
  {Johnston}, {Jorgensen}, {Kent}, {Kleinman}, {Knapp}, {Kniazev}, {Kron},
  {Krzesinski}, {Kunszt}, {Kuropatkin}, {Lamb}, {Lampeitl}, {Lee}, {Leger},
  {Li}, {Lin}, {Loh}, {Long}, {Loveday}, {Lupton}, {Malik}, {Margon},
  {Matsubara}, {McGehee}, {McKay}, {Meiksin}, {Munn}, {Nakajima}, {Nash},
  {Neilsen}, {Newberg}, {Newman}, {Nichol}, {Nicinski}, {Nieto-Santisteban},
  {Nitta}, {Okamura}, {O'Mullane}, {Ostriker}, {Owen}, {Padmanabhan},
  {Peoples}, {Pier}, {Pope}, {Quinn}, {Richards}, {Richmond}, {Rix}, {Rockosi},
  {Schlegel}, {Schneider}, {Scranton}, {Sekiguchi}, {Seljak}, {Sergey},
  {Sesar}, {Sheldon}, {Shimasaku}, {Siegmund}, {Silvestri}, {Smith}, {Smol{\v
  c}i{\'c}}, {Snedden}, {Stebbins}, {Stoughton}, {Strauss}, {SubbaRao},
  {Szalay}, {Szapudi}, {Szkody}, {Szokoly}, {Tegmark}, {Teodoro}, {Thakar},
  {Tremonti}, {Tucker}, {Uomoto}, {Vanden Berk}, {Vandenberg}, {Vogeley},
  {Voges}, {Vogt}, {Walkowicz}, {Wang}, {Weinberg}, {West}, {White}, {Wilhite},
  {Xu}, {Yanny}, {Yasuda}, {Yip}, {Yocum}, {York}, {Zehavi}, {Zibetti}, \&
  {Zucker}}]{Abazajian2004AJ....128..502A}
{Abazajian} K. {et~al.}, 2004, \aj, 128, 502

\bibitem[{{Anderson} {et~al}\mbox{.}(2001){Anderson}, {Fan}, {Richards},
  {Schneider}, {Strauss}, {Vanden Berk}, {Gunn}, {Knapp}, {Schlegel}, {Voges},
  {Yanny}, {Bahcall}, {Bernardi}, {Brinkmann}, {Brunner}, {Csab{\'a}i}, {Doi},
  {Fukugita}, {Hennessy}, {Ivezi{\'c}}, {Kunszt}, {Lamb}, {Loveday}, {Lupton},
  {McKay}, {Munn}, {Nichol}, {Szokoly}, \&
  {York}}]{Anderson2001AJ....122..503A}
{Anderson} S.~F. {et~al.}, 2001, \aj, 122, 503

\bibitem[{{Anderson} {et~al}\mbox{.}(2007){Anderson}, {Margon}, {Voges},
  {Plotkin}, {Syphers}, {Haggard}, {Collinge}, {Meyer}, {Strauss},
  {Ag{\"u}eros}, {Hall}, {Homer}, {Ivezi{\'c}}, {Richards}, {Richmond},
  {Schneider}, {Stinson}, {Vanden Berk}, \&
  {York}}]{Anderson2007AJ....133..313A}
{Anderson} S.~F. {et~al.}, 2007, \aj, 133, 313

\bibitem[{{Angel} \& {Stockman}(1980)}]{Angel1980ARA&A..18..321A}
{Angel} J.~R.~P., {Stockman} H.~S., 1980, \araa, 18, 321

\bibitem[{{Antonucci}(2012)}]{Antonucci2012A&AT...27..557A}
{Antonucci} R., 2012, Astronomical and Astrophysical Transactions, 27, 557

\bibitem[{{Bachev} {et~al}\mbox{.}(2005){Bachev}, {Strigachev}, \&
  {Semkov}}]{Bachev2005MNRAS.358..774B}
{Bachev} R., {Strigachev} A., {Semkov} E., 2005, \mnras, 358, 774

\bibitem[{{Barvainis} {et~al}\mbox{.}(2005){Barvainis}, {Leh{\'a}r},
  {Birkinshaw}, {Falcke}, \& {Blundell}}]{Barvainis2005ApJ...618..108B}
{Barvainis} R., {Leh{\'a}r} J., {Birkinshaw} M., {Falcke} H., {Blundell} K.~M.,
  2005, \apj, 618, 108

\bibitem[{{Bauer} {et~al}\mbox{.}(2009){Bauer}, {Baltay}, {Coppi}, {Donalek},
  {Drake}, {Djorgovski}, {Ellman}, {Glikman}, {Graham}, {Jerke}, {Mahabal},
  {Rabinowitz}, {Scalzo}, \& {Williams}}]{Bauer2009ApJ...705...46B}
{Bauer} A. {et~al.}, 2009, \apj, 705, 46

\bibitem[{{Becker} {et~al}\mbox{.}(1995){Becker}, {White}, \&
  {Helfand}}]{Becker1995ApJ...450..559B}
{Becker} R.~H., {White} R.~L., {Helfand} D.~J., 1995, \apj, 450, 559

\bibitem[{{Begelman} {et~al}\mbox{.}(1984){Begelman}, {Blandford}, \&
  {Rees}}]{Begelman1984RvMP...56..255B}
{Begelman} M.~C., {Blandford} R.~D., {Rees} M.~J., 1984, Reviews of Modern
  Physics, 56, 255

\bibitem[{{Blandford} \& {Rees}(1978)}]{Blandford1978PhyS...17..265B}
{Blandford} R.~D., {Rees} M.~J., 1978, PhyS, 17, 265

\bibitem[{{Blundell} \& {Beasley}(1998)}]{Blundell1998MNRAS.299..165B}
{Blundell} K.~M., {Beasley} A.~J., 1998, \mnras, 299, 165

\bibitem[{{Blundell} {et~al}\mbox{.}(2003){Blundell}, {Beasley}, \&
  {Bicknell}}]{Blundell2003ApJ...591L.103B}
{Blundell} K.~M., {Beasley} A.~J., {Bicknell} G.~V., 2003, \apjl, 591, L103

\bibitem[{{Borra} \& {Corriveau}(1984)}]{Borra1984ApJ...276..449B}
{Borra} E.~F., {Corriveau} G., 1984, \apj, 276, 449

\bibitem[{{Boyle} {et~al}\mbox{.}(2000){Boyle}, {Shanks}, {Croom}, {Smith},
  {Miller}, {Loaring}, \& {Heymans}}]{Boyle2000MNRAS.317.1014B}
{Boyle} B.~J., {Shanks} T., {Croom} S.~M., {Smith} R.~J., {Miller} L.,
  {Loaring} N., {Heymans} C., 2000, \mnras, 317, 1014

\bibitem[{{Brunthaler} {et~al}\mbox{.}(2005){Brunthaler}, {Falcke}, {Bower},
  {Aller}, {Aller}, \& {Ter{\"a}sranta}}]{Brunthaler2005A&A...435..497B}
{Brunthaler} A., {Falcke} H., {Bower} G.~C., {Aller} M.~F., {Aller} H.~D.,
  {Ter{\"a}sranta} H., 2005, \aap, 435, 497

\bibitem[{{Brunthaler} {et~al}\mbox{.}(2000){Brunthaler}, {Falcke}, {Bower},
  {Aller}, {Aller}, {Ter{\"a}sranta}, {Lobanov}, {Krichbaum}, \&
  {Patnaik}}]{Brunthaler2000A&A...357L..45B}
{Brunthaler} A. {et~al.}, 2000, \aap, 357, L45

\bibitem[{{Caccianiga} {et~al}\mbox{.}(2001){Caccianiga}, {March{\~a}},
  {Thean}, \& {Dennett-Thorpe}}]{Caccianiga2001MNRAS.328..867C}
{Caccianiga} A., {March{\~a}} M.~J.~M., {Thean} A., {Dennett-Thorpe} J., 2001,
  \mnras, 328, 867

\bibitem[{{Caplar} {et~al}\mbox{.}(2017){Caplar}, {Lilly}, \&
  {Trakhtenbrot}}]{Caplar2017ApJ...834..111C}
{Caplar} N., {Lilly} S.~J., {Trakhtenbrot} B., 2017, \apj, 834, 111

\bibitem[{{Carini} {et~al}\mbox{.}(1992){Carini}, {Miller}, {Noble}, \&
  {Goodrich}}]{Carini1992AJ....104...15C}
{Carini} M.~T., {Miller} H.~R., {Noble} J.~C., {Goodrich} B.~D., 1992, \aj,
  104, 15

\bibitem[{{Carini} {et~al}\mbox{.}(1991){Carini}, {Miller}, {Noble}, \&
  {Sadun}}]{Carini1991AJ....101.1196C}
{Carini} M.~T., {Miller} H.~R., {Noble} J.~C., {Sadun} A.~C., 1991, \aj, 101,
  1196

\bibitem[{{Cellone} {et~al}\mbox{.}(2000){Cellone}, {Romero}, \&
  {Combi}}]{Cellone2000AJ....119.1534C}
{Cellone} S.~A., {Romero} G.~E., {Combi} J.~A., 2000, \aj, 119, 1534

\bibitem[{{Chanan} {et~al}\mbox{.}(1982){Chanan}, {Downes}, {Chance}, {Margon},
  \& {Helfand}}]{Chanan1982ApJ...261L..31C}
{Chanan} G.~A., {Downes} R.~A., {Chance} D., {Margon} B., {Helfand} D.~J.,
  1982, \apjl, 261, L31

\bibitem[{{Chand} {et~al}\mbox{.}(2014){Chand}, {Kumar}, \&
  {Gopal-Krishna}}]{Chand2014MNRAS.441..726C}
{Chand} H., {Kumar} P., {Gopal-Krishna}, 2014, \mnras, 441, 726

\bibitem[{{Collier} {et~al}\mbox{.}(2001){Collier}, {Crenshaw}, {Peterson},
  {Brandt}, {Clavel}, {Edelson}, {George}, {Horne}, {Kriss}, {Mathur},
  {Netzer}, {O'Brien}, {Pogge}, {Pounds}, {Romano}, {Shemmer}, {Turner}, \&
  {Wamsteker}}]{Collier2001ApJ...561..146C}
{Collier} S. {et~al.}, 2001, \apj, 561, 146

\bibitem[{{Collinge} {et~al}\mbox{.}(2005){Collinge}, {Strauss}, {Hall},
  {Ivezi{\'c}}, {Munn}, {Schlegel}, {Zakamska}, {Anderson}, {Harris},
  {Richards}, {Schneider}, {Voges}, {York}, {Margon}, \&
  {Brinkmann}}]{Collinge2005AJ....129.2542C}
{Collinge} M.~J. {et~al.}, 2005, \aj, 129, 2542

\bibitem[{{Croom} {et~al}\mbox{.}(2001){Croom}, {Smith}, {Boyle}, {Shanks},
  {Loaring}, \& {Miller}}]{Croom2001AAONw..97....3C}
{Croom} S., {Smith} R., {Boyle} B., {Shanks} T., {Loaring} N., {Miller} L.,
  2001, Anglo-Australian Observatory Epping Newsletter, 97, 3

\bibitem[{{de Diego}(2010)}]{Diego2010AJ....139.1269D}
{de Diego} J.~A., 2010, \aj, 139, 1269

\bibitem[{{Diamond-Stanic} {et~al}\mbox{.}(2009){Diamond-Stanic}, {Fan},
  {Brandt}, {Shemmer}, {Strauss}, {Anderson}, {Carilli}, {Gibson}, {Jiang},
  {Kim}, {Richards}, {Schmidt}, {Schneider}, {Shen}, {Smith}, {Vestergaard}, \&
  {Young}}]{DiamondStanic2009ApJ...699..782D}
{Diamond-Stanic} A.~M. {et~al.}, 2009, \apj, 699, 782

\bibitem[{{Edelson} {et~al}\mbox{.}(1996){Edelson}, {Alexander}, {Crenshaw},
  {Kaspi}, {Malkan}, {Peterson}, {Warwick}, {Clavel}, {Filippenko}, {Horne},
  {Korista}, {Kriss}, {Krolik}, {Maoz}, {Nandra}, {O'Brien}, {Penton},
  {Yaqoob}, {Albrecht}, {Alloin}, {Ayres}, {Balonek}, {Barr}, {Barth},
  {Bertram}, {Bromage}, {Carini}, {Carone}, {Cheng}, {Chuvaev}, {Dietrich},
  {Dultzin-Hacyan}, {Gaskell}, {Glass}, {Goad}, {Hemar}, {Ho}, {Huchra},
  {Hutchings}, {Johnson}, {Kazanas}, {Kollatschny}, {Koratkar}, {Kovo}, {Laor},
  {MacAlpine}, {Magdziarz}, {Martin}, {Matheson}, {McCollum}, {Miller},
  {Morris}, {Oknyanskij}, {Penfold}, {Perez}, {Perola}, {Pike}, {Pogge},
  {Ptak}, {Qian}, {Recondo-Gonzalez}, {Reichert}, {Rodriguez-Espinoza},
  {Rodriguez-Pascual}, {Rokaki}, {Roland}, {Sadun}, {Salamanca}, {Santos-Lleo},
  {Shields}, {Shull}, {Smith}, {Smith}, {Snijders}, {Stirpe}, {Stoner}, {Sun},
  {Ulrich}, {van Groningen}, {Wagner}, {Wagner}, {Wanders}, {Welsh}, {Weymann},
  {Wilkes}, {Wu}, {Wurster}, {Xue}, {Zdziarski}, {Zheng}, \&
  {Zou}}]{Edelson1996ApJ...470..364E}
{Edelson} R.~A. {et~al.}, 1996, \apj, 470, 364

\bibitem[{{Falcke} {et~al}\mbox{.}(1995){Falcke}, {Malkan}, \&
  {Biermann}}]{Falcke1995A&A...298..375F}
{Falcke} H., {Malkan} M.~A., {Biermann} P.~L., 1995, \aap, 298, 375

\bibitem[{{Falcke} {et~al}\mbox{.}(1996){Falcke}, {Patnaik}, \&
  {Sherwood}}]{Falcke1996ApJ...473L..13F}
{Falcke} H., {Patnaik} A.~R., {Sherwood} W., 1996, \apjl, 473, L13

\bibitem[{{Fan} {et~al}\mbox{.}(1999){Fan}, {Strauss}, {Gunn}, {Lupton},
  {Carilli}, {Rupen}, {Schmidt}, {Moustakas}, {Davis}, {Annis}, {Bahcall},
  {Brinkmann}, {Brunner}, {Csabai}, {Doi}, {Fukugita}, {Heckman}, {Hennessy},
  {Hindsley}, {Ivezi{\'c} }, {Knapp}, {Lamb}, {Munn}, {Pauls}, {Pier},
  {Rockosi}, {Schneider}, {Szalay}, {Tucker}, \&
  {York}}]{Fan1999ApJ...526L..57F}
{Fan} X. {et~al.}, 1999, \apjl, 526, L57

\bibitem[{{Fan} {et~al}\mbox{.}(2006){Fan}, {Strauss}, {Richards}, {Hennawi},
  {Becker}, {White}, {Diamond-Stanic}, {Donley}, {Jiang}, {Kim}, {Vestergaard},
  {Young}, {Gunn}, {Lupton}, {Knapp}, {Schneider}, {Brandt}, {Bahcall},
  {Barentine}, {Brinkmann}, {Brewington}, {Fukugita}, {Harvanek}, {Kleinman},
  {Krzesinski}, {Long}, {Neilsen}, {Nitta}, {Snedden}, \&
  {Voges}}]{Fan2006AJ....131.1203F}
{Fan} X. {et~al.}, 2006, \aj, 131, 1203

\bibitem[{{Fugmann}(1988)}]{Fugmann1988A&A...205...86F}
{Fugmann} W., 1988, \aap, 205, 86

\bibitem[{{Gallo}(2006)}]{Gallo2006MNRAS.365..960G}
{Gallo} L.~C., 2006, \mnras, 365, 960

\bibitem[{{Garcia} {et~al}\mbox{.}(1999){Garcia}, {Sodr{\'e}}, {Jablonski}, \&
  {Terlevich}}]{1999MNRAS.309..803G}
{Garcia} A., {Sodr{\'e}} L., {Jablonski} F.~J., {Terlevich} R.~J., 1999,
  \mnras, 309, 803

\bibitem[{{Gopal-Krishna} {et~al}\mbox{.}(2013){Gopal-Krishna}, {Joshi}, \&
  {Chand}}]{Gopal2013MNRAS.430.1302G}
{Gopal-Krishna}, {Joshi} R., {Chand} H., 2013, \mnras, 430, 1302

\bibitem[{{Gopal-Krishna} {et~al}\mbox{.}(1995){Gopal-Krishna}, {Sagar}, \&
  {Wiita}}]{1995MNRAS.274..701G}
{Gopal-Krishna}, {Sagar} R., {Wiita} P.~J., 1995, \mnras, 274, 701

\bibitem[{{Goyal} {et~al}\mbox{.}(2010){Goyal}, {Gopal-Krishna}, {Joshi},
  {Sagar}, {Wiita}, {Anupama}, \& {Sahu}}]{Goyal2010MNRAS.401.2622G}
{Goyal} A., {Gopal-Krishna}, {Joshi} S., {Sagar} R., {Wiita} P.~J., {Anupama}
  G.~C., {Sahu} D.~K., 2010, \mnras, 401, 2622

\bibitem[{{Goyal} {et~al}\mbox{.}(2012){Goyal}, {Gopal-Krishna}, {Wiita},
  {Anupama}, {Sahu}, {Sagar}, \& {Joshi}}]{Goyal2012A&A...544A..37G}
{Goyal} A., {Gopal-Krishna}, {Wiita} P.~J., {Anupama} G.~C., {Sahu} D.~K.,
  {Sagar} R., {Joshi} S., 2012, \aap, 544, A37

\bibitem[{{Goyal} {et~al}\mbox{.}(2013){Goyal}, {Gopal-Krishna}, {Wiita},
  {Stalin}, \& {Sagar}}]{Goyal2013MNRAS.435.1300G}
{Goyal} A., {Gopal-Krishna}, {Wiita} P.~J., {Stalin} C.~S., {Sagar} R., 2013,
  \mnras, 435, 1300(GGWSS13)

\bibitem[{{Heidt} \& {Nilsson}(2011)}]{Heidt2011A&A...529A.162H}
{Heidt} J., {Nilsson} K., 2011, \aap, 529, A162

\bibitem[{{Herrera Ruiz} {et~al}\mbox{.}(2016){Herrera Ruiz}, {Middelberg},
  {Norris}, \& {Maini}}]{Herrera2016A&A...589L...2H}
{Herrera Ruiz} N., {Middelberg} E., {Norris} R.~P., {Maini} A., 2016, \aap,
  589, L2

\bibitem[{{Hewett} \& {Wild}(2010)}]{Hewett2010MNRAS.405.2302H}
{Hewett} P.~C., {Wild} V., 2010, \mnras, 405, 2302

\bibitem[{{Hryniewicz} {et~al}\mbox{.}(2010){Hryniewicz}, {Czerny},
  {Niko{\l}ajuk}, \& {Kuraszkiewicz}}]{Hryniewicz2010MNRAS.404.2028H}
{Hryniewicz} K., {Czerny} B., {Niko{\l}ajuk} M., {Kuraszkiewicz} J., 2010,
  \mnras, 404, 2028

\bibitem[{{Impey} \& {Brand}(1982)}]{Impey1982MNRAS.201..849I}
{Impey} C.~D., {Brand} P.~W.~J.~L., 1982, \mnras, 201, 849

\bibitem[{{Jang} \& {Miller}(1997)}]{1997AJ....114..565J}
{Jang} M., {Miller} H.~R., 1997, \aj, 114, 565

\bibitem[{{Jannuzi} {et~al}\mbox{.}(1993){Jannuzi}, {Green}, \&
  {French}}]{Jannuzi1993ApJ...404..100J}
{Jannuzi} B.~T., {Green} R.~F., {French} H., 1993, \apj, 404, 100

\bibitem[{{Jannuzi} {et~al}\mbox{.}(1994){Jannuzi}, {Smith}, \&
  {Elston}}]{Jannuzi1994ApJ...428..130J}
{Jannuzi} B.~T., {Smith} P.~S., {Elston} R., 1994, \apj, 428, 130

\bibitem[{{Joshi} {et~al}\mbox{.}(2011){Joshi}, {Chand}, {Gupta}, \&
  {Wiita}}]{Joshi2011MNRAS.412.2717J}
{Joshi} R., {Chand} H., {Gupta} A.~C., {Wiita} P.~J., 2011, \mnras, 412, 2717

\bibitem[{{Kellermann} {et~al}\mbox{.}(1989){Kellermann}, {Sramek}, {Schmidt},
  {Shaffer}, \& {Green}}]{Kellermann1989AJ.....98.1195K}
{Kellermann} K.~I., {Sramek} R., {Schmidt} M., {Shaffer} D.~B., {Green} R.,
  1989, \aj, 98, 1195

\bibitem[{{K{\"u}gler} {et~al}\mbox{.}(2015){K{\"u}gler}, {Gianniotis}, \&
  {Polsterer}}]{Kugler2015MNRAS.451.3385K}
{K{\"u}gler} S.~D., {Gianniotis} N., {Polsterer} K.~L., 2015, \mnras, 451, 3385

\bibitem[{{K{\"u}gler} {et~al}\mbox{.}(2014{\natexlab{a}}){K{\"u}gler},
  {Nilsson}, {Heidt}, {Esser}, \& {Schultz}}]{Kuegler2014A&A...569A..95K}
{K{\"u}gler} S.~D., {Nilsson} K., {Heidt} J., {Esser} J., {Schultz} T.,
  2014{\natexlab{a}}, \aap, 569, A95

\bibitem[{{K{\"u}gler} {et~al}\mbox{.}(2014{\natexlab{b}}){K{\"u}gler},
  {Nilsson}, {Heidt}, {Esser}, \& {Schultz}}]{Kugler2014A&A...569A..95K}
{K{\"u}gler} S.~D., {Nilsson} K., {Heidt} J., {Esser} J., {Schultz} T.,
  2014{\natexlab{b}}, \aap, 569, A95

\bibitem[{{Kumar} {et~al}\mbox{.}(2016){Kumar}, {Chand}, \&
  {Gopal-Krishna}}]{Kumar2016MNRAS.461..666K}
{Kumar} P., {Chand} H., {Gopal-Krishna}, 2016, \mnras, 461, 666

\bibitem[{{Kumar} {et~al}\mbox{.}(2015){Kumar}, {Gopal-Krishna}, \&
  {Chand}}]{Kumar2015MNRAS.448.1463K}
{Kumar} P., {Gopal-Krishna}, {Chand} H., 2015, \mnras, 448, 1463

\bibitem[{{Lane} {et~al}\mbox{.}(2011){Lane}, {Shemmer}, {Diamond-Stanic},
  {Fan}, {Anderson}, {Brandt}, {Plotkin}, {Richards}, {Schneider}, \&
  {Strauss}}]{Lane2011ApJ...743..163L}
{Lane} R.~A. {et~al.}, 2011, \apj, 743, 163

\bibitem[{{Laor} \& {Davis}(2011)}]{Laor2011MNRAS.417..681L}
{Laor} A., {Davis} S.~W., 2011, \mnras, 417, 681

\bibitem[{{Leighly} {et~al}\mbox{.}(2007){Leighly}, {Halpern}, {Jenkins},
  {Grupe}, {Choi}, \& {Prescott}}]{Leighly2007ApJ...663..103L}
{Leighly} K.~M., {Halpern} J.~P., {Jenkins} E.~B., {Grupe} D., {Choi} J.,
  {Prescott} K.~B., 2007, \apj, 663, 103

\bibitem[{{Leipski} {et~al}\mbox{.}(2006){Leipski}, {Falcke}, {Bennert}, \&
  {H{\"u}ttemeister}}]{Leipski2006A&A...455..161L}
{Leipski} C., {Falcke} H., {Bennert} N., {H{\"u}ttemeister} S., 2006, \aap,
  455, 161

\bibitem[{{Liao} {et~al}\mbox{.}(2016){Liao}, {Xin}, {Fan}, {Weng}, {Li},
  {Chen}, \& {Fan}}]{Liao2016ApJS..226...17L}
{Liao} N.-H., {Xin} Y.-L., {Fan} X.-L., {Weng} S.-S., {Li} S.-K., {Chen} L.,
  {Fan} Y.-Z., 2016, \apjs, 226, 17

\bibitem[{{Liu} {et~al}\mbox{.}(2016){Liu}, {Li}, {Gu}, \&
  {Guo}}]{Liu2016MNRAS.462L..56L}
{Liu} H., {Li} S.-L., {Gu} M., {Guo} H., 2016, \mnras, 462, L56

\bibitem[{{Londish} {et~al}\mbox{.}(2002){Londish}, {Croom}, {Boyle}, {Shanks},
  {Outram}, {Sadler}, {Loaring}, {Smith}, {Miller}, \&
  {Maxted}}]{Londish2002MNRAS.334..941L}
{Londish} D. {et~al.}, 2002, \mnras, 334, 941

\bibitem[{{Londish} {et~al}\mbox{.}(2007){Londish}, {Croom}, {Heidt}, {Boyle},
  {Sadler}, {Whiting}, {Rector}, {Pursimo}, \&
  {Chynoweth}}]{Londish2007MNRAS.374..556L}
{Londish} D. {et~al.}, 2007, \mnras, 374, 556

\bibitem[{{Luo} {et~al}\mbox{.}(2015){Luo}, {Brandt}, {Hall}, {Wu}, {Anderson},
  {Garmire}, {Gibson}, {Plotkin}, {Richards}, {Schneider}, {Shemmer}, \&
  {Shen}}]{Luo2015ApJ...805..122L}
{Luo} B. {et~al.}, 2015, \apj, 805, 122

\bibitem[{{McDowell} {et~al}\mbox{.}(1995){McDowell}, {Canizares}, {Elvis},
  {Lawrence}, {Markoff}, {Mathur}, \& {Wilkes}}]{McDowell1995ApJ...450..585M}
{McDowell} J.~C., {Canizares} C., {Elvis} M., {Lawrence} A., {Markoff} S.,
  {Mathur} S., {Wilkes} B.~J., 1995, \apj, 450, 585

\bibitem[{{Meusinger} \& {Balafkan}(2014)}]{Meusinger2014A&A...568A.114M}
{Meusinger} H., {Balafkan} N., 2014, \aap, 568, A114

\bibitem[{{Miller} {et~al}\mbox{.}(1993){Miller}, {Rawlings}, \&
  {Saunders}}]{Miller1993MNRAS.263..425M}
{Miller} P., {Rawlings} S., {Saunders} R., 1993, \mnras, 263, 425

\bibitem[{{Monet} {et~al}\mbox{.}(2003){Monet}, {Levine}, {Canzian}, {Ables},
  {Bird}, {Dahn}, {Guetter}, {Harris}, {Henden}, {Leggett}, {Levison},
  {Luginbuhl}, {Martini}, {Monet}, {Munn}, {Pier}, {Rhodes}, {Riepe}, {Sell},
  {Stone}, {Vrba}, {Walker}, {Westerhout}, {Brucato}, {Reid}, {Schoening},
  {Hartley}, {Read}, \& {Tritton}}]{Monet2003AJ....125..984M}
{Monet} D.~G. {et~al.}, 2003, \aj, 125, 984

\bibitem[{{Murray} \& {Chiang}(1997)}]{Murray1997ApJ...474...91M}
{Murray} N., {Chiang} J., 1997, \apj, 474, 91

\bibitem[{{Nicastro} {et~al}\mbox{.}(2003){Nicastro}, {Martocchia}, \&
  {Matt}}]{Nicastro2003ApJ...589L..13N}
{Nicastro} F., {Martocchia} A., {Matt} G., 2003, \apjl, 589, L13

\bibitem[{{Niko{\l}ajuk} \& {Walter}(2012)}]{Nikolajuk2012MNRAS.420.2518N}
{Niko{\l}ajuk} M., {Walter} R., 2012, \mnras, 420, 2518

\bibitem[{{Padovani} {et~al}\mbox{.}(2007){Padovani}, {Giommi}, {Landt}, \&
  {Perlman}}]{Padovani2007ApJ...662..182P}
{Padovani} P., {Giommi} P., {Landt} H., {Perlman} E.~S., 2007, \apj, 662, 182

\bibitem[{{Paltani} {et~al}\mbox{.}(1998){Paltani}, {Courvoisier}, \&
  {Walter}}]{Paltani1998A&A...340...47P}
{Paltani} S., {Courvoisier} T.~J.-L., {Walter} R., 1998, \aap, 340, 47

\bibitem[{{Perlman} {et~al}\mbox{.}(1996){Perlman}, {Stocke}, {Schachter},
  {Elvis}, {Ellingson}, {Urry}, {Potter}, {Impey}, \&
  {Kolchinsky}}]{Perlman1996ApJS..104..251P}
{Perlman} E.~S. {et~al.}, 1996, \apjs, 104, 251

\bibitem[{{Plotkin} {et~al}\mbox{.}(2010{\natexlab{a}}){Plotkin}, {Anderson},
  {Brandt}, {Diamond-Stanic}, {Fan}, {Hall}, {Kimball}, {Richmond},
  {Schneider}, {Shemmer}, {Voges}, {York}, {Bahcall}, {Snedden}, {Bizyaev},
  {Brewington}, {Malanushenko}, {Malanushenko}, {Oravetz}, {Pan}, \&
  {Simmons}}]{Plotkin2010AJ....139..390P}
{Plotkin} R.~M. {et~al.}, 2010{\natexlab{a}}, \aj, 139, 390

\bibitem[{{Plotkin} {et~al}\mbox{.}(2010{\natexlab{b}}){Plotkin}, {Anderson},
  {Brandt}, {Diamond-Stanic}, {Fan}, {MacLeod}, {Schneider}, \&
  {Shemmer}}]{Plotkin2010ApJ...721..562P}
{Plotkin} R.~M., {Anderson} S.~F., {Brandt} W.~N., {Diamond-Stanic} A.~M.,
  {Fan} X., {MacLeod} C.~L., {Schneider} D.~P., {Shemmer} O.,
  2010{\natexlab{b}}, \apj, 721, 562

\bibitem[{{Plotkin} {et~al}\mbox{.}(2008){Plotkin}, {Anderson}, {Hall},
  {Margon}, {Voges}, {Schneider}, {Stinson}, \&
  {York}}]{Plotkin2008AJ....135.2453P}
{Plotkin} R.~M., {Anderson} S.~F., {Hall} P.~B., {Margon} B., {Voges} W.,
  {Schneider} D.~P., {Stinson} G., {York} D.~G., 2008, \aj, 135, 2453

\bibitem[{{Plotkin} {et~al}\mbox{.}(2015){Plotkin}, {Shemmer}, {Trakhtenbrot},
  {Anderson}, {Brandt}, {Fan}, {Gallo}, {Lira}, {Luo}, {Richards}, {Schneider},
  {Strauss}, \& {Wu}}]{Plotkin2015ApJ...805..123P}
{Plotkin} R.~M. {et~al.}, 2015, \apj, 805, 123

\bibitem[{{Prabhu} \& {Anupama}(2010)}]{Prabhu2010ASInC...1..193P}
{Prabhu} T.~P., {Anupama} G.~C., 2010, in Astronomical Society of India
  Conference Series, Vol.~1, Astronomical Society of India Conference Series

\bibitem[{{Richards} {et~al}\mbox{.}(2009){Richards}, {Myers}, {Gray},
  {Riegel}, {Nichol}, {Brunner}, {Szalay}, {Schneider}, \&
  {Anderson}}]{Richards2009ApJS..180...67R}
{Richards} G.~T. {et~al.}, 2009, \apjs, 180, 67

\bibitem[{{Rokaki} {et~al}\mbox{.}(1993){Rokaki}, {Collin-Souffrin}, \&
  {Magnan}}]{Rokaki1993A&A...272....8R}
{Rokaki} E., {Collin-Souffrin} S., {Magnan} C., 1993, \aap, 272, 8

\bibitem[{{Ruan} {et~al}\mbox{.}(2012){Ruan}, {Anderson}, {MacLeod}, {Becker},
  {Burnett}, {Davenport}, {Ivezi{\'c}}, {Kochanek}, {Plotkin}, {Sesar}, \&
  {Stuart}}]{Ruan2012ApJ...760...51R}
{Ruan} J.~J. {et~al.}, 2012, \apj, 760, 51

\bibitem[{{Sagar} {et~al}\mbox{.}(2011){Sagar}, {Omar}, {Kumar}, {Gopinathan},
  {Pandey}, {Bangia}, {Pant}, {Shukla}, \&
  {Yadava}}]{Sagar2011Csi...101...8.25}
{Sagar} R. {et~al.}, 2011, CURRENT-SCIENCE, 101, 8

\bibitem[{{Sagar} {et~al}\mbox{.}(2004){Sagar}, {Stalin}, {Gopal-Krishna}, \&
  {Wiita}}]{Sagar2004MNRAS.348..176S}
{Sagar} R., {Stalin} C.~S., {Gopal-Krishna}, {Wiita} P.~J., 2004, \mnras, 348,
  176

\bibitem[{{Schneider} {et~al}\mbox{.}(2005){Schneider}, {Hall}, {Richards},
  {Vanden Berk}, {Anderson}, {Fan}, {Jester}, {Stoughton}, {Strauss},
  {SubbaRao}, {Brandt}, {Gunn}, {Yanny}, {Bahcall}, {Barentine}, {Blanton},
  {Boroski}, {Brewington}, {Brinkmann}, {Brunner}, {Csabai}, {Doi},
  {Eisenstein}, {Frieman}, {Fukugita}, {Gray}, {Harvanek}, {Heckman},
  {Ivezi{\'c}}, {Kent}, {Kleinman}, {Knapp}, {Kron}, {Krzesinski}, {Long},
  {Loveday}, {Lupton}, {Margon}, {Munn}, {Neilsen}, {Newberg}, {Newman},
  {Nichol}, {Nitta}, {Pier}, {Rockosi}, {Saxe}, {Schlegel}, {Snedden},
  {Szalay}, {Thakar}, {Uomoto}, {Voges}, \&
  {York}}]{Schneider2005AJ....130..367S}
{Schneider} D.~P. {et~al.}, 2005, \aj, 130, 367

\bibitem[{{Shemmer} {et~al}\mbox{.}(2009){Shemmer}, {Brandt}, {Anderson},
  {Diamond-Stanic}, {Fan}, {Richards}, {Schneider}, \&
  {Strauss}}]{Shemmer2009ApJ...696..580S}
{Shemmer} O., {Brandt} W.~N., {Anderson} S.~F., {Diamond-Stanic} A.~M., {Fan}
  X., {Richards} G.~T., {Schneider} D.~P., {Strauss} M.~A., 2009, \apj, 696,
  580

\bibitem[{{Shemmer} {et~al}\mbox{.}(2006){Shemmer}, {Brandt}, {Schneider},
  {Fan}, {Strauss}, {Diamond-Stanic}, {Richards}, {Anderson}, {Gunn}, \&
  {Brinkmann}}]{Shemmer2006ApJ...644...86S}
{Shemmer} O. {et~al.}, 2006, \apj, 644, 86

\bibitem[{{Shemmer} {et~al}\mbox{.}(2010){Shemmer}, {Trakhtenbrot}, {Anderson},
  {Brandt}, {Diamond-Stanic}, {Fan}, {Lira}, {Netzer}, {Plotkin}, {Richards},
  {Schneider}, \& {Strauss}}]{Shemmer2010ApJ...722L.152S}
{Shemmer} O. {et~al.}, 2010, \apjl, 722, L152

\bibitem[{{Smith} {et~al}\mbox{.}(2007){Smith}, {Williams}, {Schmidt},
  {Diamond-Stanic}, \& {Means}}]{Smith2007ApJ...663..118S}
{Smith} P.~S., {Williams} G.~G., {Schmidt} G.~D., {Diamond-Stanic} A.~M.,
  {Means} D.~L., 2007, \apj, 663, 118

\bibitem[{{Stalin} {et~al}\mbox{.}(2004{\natexlab{a}}){Stalin},
  {Gopal-Krishna}, {Sagar}, \& {Wiita}}]{Stalin2004MNRAS.350..175S}
{Stalin} C.~S., {Gopal-Krishna}, {Sagar} R., {Wiita} P.~J., 2004{\natexlab{a}},
  \mnras, 350, 175

\bibitem[{{Stalin} {et~al}\mbox{.}(2004{\natexlab{b}}){Stalin}, {Gopal
  Krishna}, {Sagar}, \& {Wiita}}]{Stalin2004JApA...25....1S}
{Stalin} C.~S., {Gopal Krishna}, {Sagar} R., {Wiita} P.~J., 2004{\natexlab{b}},
  Journal of Astrophysics and Astronomy, 25, 1

\bibitem[{{Stalin} \& {Srianand}(2005)}]{Stalin2005MNRAS.359.1022S}
{Stalin} C.~S., {Srianand} R., 2005, \mnras, 359, 1022

\bibitem[{{Stetson}(1987)}]{1987PASP...99..191S}
{Stetson} P.~B., 1987, \pasp, 99, 191

\bibitem[{{Stetson}(1992)}]{1992ASPC...25..297S}
{Stetson} P.~B., 1992, in Astronomical Society of the Pacific Conference
  Series, Vol.~25, Astronomical Data Analysis Software and Systems I, {Worrall}
  D.~M., {Biemesderfer} C., {Barnes} J., eds., p. 297

\bibitem[{{Stickel} {et~al}\mbox{.}(1991){Stickel}, {Padovani}, {Urry},
  {Fried}, \& {Kuehr}}]{1991ApJ...374..431S}
{Stickel} M., {Padovani} P., {Urry} C.~M., {Fried} J.~W., {Kuehr} H., 1991,
  \apj, 374, 431

\bibitem[{{Stocke}(2001)}]{Stocke2001ASPC..227..184S}
{Stocke} J.~T., 2001, in Astronomical Society of the Pacific Conference Series,
  Vol. 227, Blazar Demographics and Physics, {Padovani} P., {Urry} C.~M., eds.,
  p. 184

\bibitem[{{Stocke} {et~al}\mbox{.}(1990){Stocke}, {Morris}, {Gioia},
  {Maccacaro}, {Schild}, \& {Wolter}}]{Stocke1990ApJ...348..141S}
{Stocke} J.~T., {Morris} S.~L., {Gioia} I., {Maccacaro} T., {Schild} R.~E.,
  {Wolter} A., 1990, \apj, 348, 141

\bibitem[{{Stocke} {et~al}\mbox{.}(1991){Stocke}, {Morris}, {Gioia},
  {Maccacaro}, {Schild}, {Wolter}, {Fleming}, \&
  {Henry}}]{Stocke1991ApJS...76..813S}
{Stocke} J.~T., {Morris} S.~L., {Gioia} I.~M., {Maccacaro} T., {Schild} R.,
  {Wolter} A., {Fleming} T.~A., {Henry} J.~P., 1991, \apjs, 76, 813

\bibitem[{{Teraesranta} {et~al}\mbox{.}(1998){Teraesranta}, {Tornikoski},
  {Mujunen}, {Karlamaa}, {Valtonen}, {Henelius}, {Urpo}, {Lainela}, {Pursimo},
  {Nilsson}, {Wiren}, {Laehteenmaeki}, {Korpi}, {Rekola}, {Heinaemaeki},
  {Hanski}, {Nurmi}, {Kokkonen}, {Keinaenen}, {Joutsamo}, {Oksanen},
  {Pietilae}, {Valtaoja}, {Valtonen}, \&
  {Koenoenen}}]{Teraesranta1998A&AS..132..305T}
{Teraesranta} H. {et~al.}, 1998, \aaps, 132, 305

\bibitem[{{Ulvestad} {et~al}\mbox{.}(2005){Ulvestad}, {Antonucci}, \&
  {Barvainis}}]{Ulvestad2005ApJ...621..123U}
{Ulvestad} J.~S., {Antonucci} R.~R.~J., {Barvainis} R., 2005, \apj, 621, 123

\bibitem[{{Urry} \& {Padovani}(1995)}]{Urry1995PASP..107..803U}
{Urry} C.~M., {Padovani} P., 1995, \pasp, 107, 803

\bibitem[{{Villforth} {et~al}\mbox{.}(2010){Villforth}, {Koekemoer}, \&
  {Grogin}}]{Villforth2010ApJ...723..737V}
{Villforth} C., {Koekemoer} A.~M., {Grogin} N.~A., 2010, \apj, 723, 737

\bibitem[{{Villforth} {et~al}\mbox{.}(2009){Villforth}, {Nilsson},
  {{\O}stensen}, {Heidt}, {Niemi}, \& {Pforr}}]{Villforth2009MNRAS.397.1893V}
{Villforth} C., {Nilsson} K., {{\O}stensen} R., {Heidt} J., {Niemi} S.-M.,
  {Pforr} J., 2009, \mnras, 397, 1893

\bibitem[{{Wang} {et~al}\mbox{.}(2006){Wang}, {Zhou}, {Wang}, {Lu}, \&
  {Lu}}]{Wang2006ApJ...645..856W}
{Wang} T.-G., {Zhou} H.-Y., {Wang} J.-X., {Lu} Y.-J., {Lu} Y., 2006, \apj, 645,
  856

\bibitem[{{Welsh} {et~al}\mbox{.}(2011){Welsh}, {Wheatley}, \&
  {Neil}}]{Welsh2011A&A...527A..15W}
{Welsh} B.~Y., {Wheatley} J.~M., {Neil} J.~D., 2011, \aap, 527, A15

\bibitem[{{Wiita}(1996)}]{Wiita1996ASPC..110...42W}
{Wiita} P.~J., 1996, in Astronomical Society of the Pacific Conference Series,
  Vol. 110, Blazar Continuum Variability, {Miller} H.~R., {Webb} J.~R., {Noble}
  J.~C., eds., p.~42

\bibitem[{{Wills} {et~al}\mbox{.}(1992){Wills}, {Wills}, {Breger}, {Antonucci},
  \& {Barvainis}}]{Wills1992ApJ...398..454W}
{Wills} B.~J., {Wills} D., {Breger} M., {Antonucci} R.~R.~J., {Barvainis} R.,
  1992, \apj, 398, 454

\bibitem[{{Wu} {et~al}\mbox{.}(2012){Wu}, {Brandt}, {Anderson},
  {Diamond-Stanic}, {Hall}, {Plotkin}, {Schneider}, \&
  {Shemmer}}]{Wu2012ApJ...747...10W}
{Wu} J., {Brandt} W.~N., {Anderson} S.~F., {Diamond-Stanic} A.~M., {Hall}
  P.~B., {Plotkin} R.~M., {Schneider} D.~P., {Shemmer} O., 2012, \apj, 747, 10

\bibitem[{{Wu} {et~al}\mbox{.}(2011){Wu}, {Brandt}, {Hall}, {Gibson},
  {Richards}, {Schneider}, {Shemmer}, {Just}, \&
  {Schmidt}}]{Wu2011ApJ...736...28W}
{Wu} J. {et~al.}, 2011, \apj, 736, 28

\bibitem[{{York} {et~al}\mbox{.}(2000){York}, {Adelman}, {Anderson},
  {Anderson}, {Annis}, {Bahcall}, {Bakken}, {Barkhouser}, {Bastian}, {Berman},
  {Boroski}, {Bracker}, {Briegel}, {Briggs}, {Brinkmann}, {Brunner}, {Burles},
  {Carey}, {Carr}, {Castander}, {Chen}, {Colestock}, {Connolly}, {Crocker},
  {Csabai}, {Czarapata}, {Davis}, {Doi}, {Dombeck}, {Eisenstein}, {Ellman},
  {Elms}, {Evans}, {Fan}, {Federwitz}, {Fiscelli}, {Friedman}, {Frieman},
  {Fukugita}, {Gillespie}, {Gunn}, {Gurbani}, {de Haas}, {Haldeman}, {Harris},
  {Hayes}, {Heckman}, {Hennessy}, {Hindsley}, {Holm}, {Holmgren}, {Huang},
  {Hull}, {Husby}, {Ichikawa}, {Ichikawa}, {Ivezi{\'c}}, {Kent}, {Kim},
  {Kinney}, {Klaene}, {Kleinman}, {Kleinman}, {Knapp}, {Korienek}, {Kron},
  {Kunszt}, {Lamb}, {Lee}, {Leger}, {Limmongkol}, {Lindenmeyer}, {Long},
  {Loomis}, {Loveday}, {Lucinio}, {Lupton}, {MacKinnon}, {Mannery}, {Mantsch},
  {Margon}, {McGehee}, {McKay}, {Meiksin}, {Merelli}, {Monet}, {Munn},
  {Narayanan}, {Nash}, {Neilsen}, {Neswold}, {Newberg}, {Nichol}, {Nicinski},
  {Nonino}, {Okada}, {Okamura}, {Ostriker}, {Owen}, {Pauls}, {Peoples},
  {Peterson}, {Petravick}, {Pier}, {Pope}, {Pordes}, {Prosapio},
  {Rechenmacher}, {Quinn}, {Richards}, {Richmond}, {Rivetta}, {Rockosi},
  {Ruthmansdorfer}, {Sandford}, {Schlegel}, {Schneider}, {Sekiguchi}, {Sergey},
  {Shimasaku}, {Siegmund}, {Smee}, {Smith}, {Snedden}, {Stone}, {Stoughton},
  {Strauss}, {Stubbs}, {SubbaRao}, {Szalay}, {Szapudi}, {Szokoly}, {Thakar},
  {Tremonti}, {Tucker}, {Uomoto}, {Vanden Berk}, {Vogeley}, {Waddell}, {Wang},
  {Watanabe}, {Weinberg}, {Yanny}, {Yasuda}, \& {SDSS
  Collaboration}}]{York2000AJ....120.1579Y}
{York} D.~G. {et~al.}, 2000, \aj, 120, 1579

\end{thebibliography}
\label{lastpage}
\end{document}